%% file: ms.tex
\documentclass[12pt,preprint]{aastex}
\usepackage{epsf}
\input{defs.tex}
\begin{document}

\title{A \pc\ Sonata III. Growth of Charon from a Circum-Pluto Ring of Debris}
\vskip 7ex
\author{Scott J. Kenyon}
\affil{Smithsonian Astrophysical Observatory,
60 Garden Street, Cambridge, MA 02138}
\email{e-mail: skenyon@cfa.harvard.edu}

\author{Benjamin C. Bromley}
\affil{Department of Physics \& Astronomy, University of Utah,
201 JFB, Salt Lake City, UT 84112}
\email{e-mail: bromley@physics.utah.edu}

\begin{abstract}

Current theory considers two options for the formation of the \pc\ binary
\citep[e.g.,][]{canup2005,canup2011,desch2015}. In the `hit-and-run' model, 
a lower mass projectile barely hits the more massive Pluto, kicks up some 
debris, and remains bound to Pluto \citep[see also][]{asphaug2006}.  In a 
`graze-and-merge' scenario, the projectile ejects substantial debris as it 
merges with Pluto \citep[see also][]{canup2001}. To investigate the 
graze-and-merge idea in more detail, we consider the growth of Charon-mass 
objects within a circum-Pluto ring of solids.  Numerical calculations 
demonstrate that Charon analogs form rapidly within a swarm of planetesimals 
with initial radii $\r0 \approx$ 145--230~km. On time scales of $\sim$ 
30--100 days, newly-formed Charon analogs have semimajor axes, $a \approx$ 
5--6~\rp, and orbital eccentricities, $e \approx$ 0.1--0.3, similar to Charon 
analogs that remain bound after hit-and-run collisions with Pluto.  Although 
the early growth of Charon analogs generates rings of small particles at 
$a \approx$ 50--275~\rp, ejection of several 145--230~km leftovers by the 
central \pc\ binary removes these small solids in 10--100~yr. Simple 
estimates suggest small particles might survive the passage of 10--20~km 
objects ejected by the central binary.  Our results indicate that the 
\pc\ circumbinary satellite system was not formed by a graze-and-merge 
impact 
when the formation of Charon within a circum-Pluto disk leads to the ejection 
of several 100--200~km particles through the orbital plane of the \pc\ binary.
If a growing Charon ejects only much smaller particles, however,
graze-and-merge impacts are a plausible formation channel for the \pc\ binary
and an ensemble of small, circumbinary satellites.

\end{abstract}


\section{INTRODUCTION}
\label{sec: intro}

Inside the protosolar nebula, Pluto and other trans-Neptunian objects (TNOs)
begin their lives as disparate small particles with a broad range in sizes 
and (probably) semimajor axes \citep[e.g.,][]{armitage2013,birn2016,lammer2018,
raymond2018}.
Within a few thousand yr, these particles grow to cm and larger sizes 
\citep[e.g.,][]{dulle2005,brauer2008,birnstiel2010,drazkowska2018,
lenz2019}.  During the next 0.1--0.3~Myr \citep[e.g.,][]{najita2014}, 
agglomeration and concentration processes generate km-sized or larger 
planetesimals \citep[e.g.,][]{youdin2005,jkh06,johan2007,lamb2012,birn2016,
arakawa2016,blum2018}.  In some theories, planetesimals become planets by 
accreting cm-sized pebbles \citep[e.g.,][and references therein]{kb2009,
bk2011a,chambers2016,johansen2017,lenz2019}. In others, successive collisional 
mergers of planetesimals and larger objects eventually produce several 
planets \citep[e.g.,][and references therein]{stcol1997a,kb2008,kb2010,raymond2011,
kb2012,hansen2012,schlicht2013}.

Numerical calculations suggest the transformation of a swarm of planetesimals 
into a planetary system is often a chaotic process, with numerous `giant impacts' 
between planet-mass objects \citep[e.g.,][]{agnor1999,chambers2001a,asphaug2006,
genda2012,chambers2013,asphaug2014,quintana2016}.  Many of these binary collisions 
simply add to the ever-growing mass of the newly-formed planet. Others may 
completely disrupt the more massive object of the pair. Sometimes, the geometry 
allows the impactor to eject a cloud of debris or to survive, leading to the 
formation of a binary planet similar to the Earth--Moon 
\citep[e.g.,][]{hartmann1975} or the \pc\ system \citep[e.g.,][]{mckinnon1989}. 

Although rocky and icy planets form in similar ways, collision outcomes depend
on the geometry and the velocity of the impact \citep[e.g.,][]{benz1999,lein2002,
asphaug2006,lein2012,emsenhuber2019,arakawa2019}. Among planetary embryos with
$a \lesssim$ 2~au and $e \gtrsim$ 0.1, collisions have typical relative velocities
$v_{rel} \gtrsim$ 3--10~\kms. Despite their large relative velocities, rocky planetary 
embryos in the terrestrial zone also have large binding energies. Thus, terrestrial
planet formation is fairly robust; Earth-mass or larger planets grow from much less
massive objects on time scales of 10--100~Myr \citep{kb2006,chambers2008,raymond2011,
hansen2012,genda2015a}. At $a \gtrsim$ 20~au, the relative velocities of icy planetary
embryos are smaller, $v_{rel} \lesssim$ 1--3~\kms. With much smaller binding energies,
however, it is much easier for collisions to prevent the growth of massive planets 
\citep[e.g.,][]{kb2008,kb2010,kb2012}. Dynamical interactions with gas giants at 
smaller $a$ can also halt growth and limit the production of Earth-mass or larger
planets beyond 20~au \citep[e.g.,][]{levison2003b,levison2008,morbi2008}.

To explore the physical conditions required for a giant impact to produce the
\pc\ system, \citet{canup2005,canup2011} performed an extensive suite of SPH 
calculations with a range of compositions, masses, spins, and impact angles
for a collision between proto-Pluto and proto-Charon.  In `graze-and-merge' 
collisions with impact velocities $v_i$ comparable to the escape velocity 
$v_{esc}$ ($v_c \approx$ 1.0--1.2 $v_{esc}$), the impactor merges with Pluto 
and ejects a massive disk of debris \citep[for other applications, see][and 
references therein]{lein2010,canup2013,nakajima2014,barr2016,barr2017}. 
Collisions within the disk eventually generate a lower density satellite with 
the mass of Charon. `Hit-and-run' collisions with $v_c \approx$ 1.0--1.2 $v_{esc}$
allow a proto-Charon to survive and remain bound to Pluto \citep[for other 
examples of hit-and-run collisions, including those where the impactor 
escapes the target, see][]{asphaug2006,chambers2013,emsenhuber2019}. Many 
hit-and-run collisions generate a disk of debris with a mass sufficient to 
produce the four known circumbinary satellites. From simulations of planet 
formation within the protosolar nebula, both types of collisions are probably 
common \citep[e.g.,][]{kb2012,kb2014b}.

Here, we describe a suite of numerical simulations designed to establish 
whether a Charon-mass satellite can form out of a disk of debris orbiting a 
Pluto-mass planet.  Although several analyses  of the \pc\ system currently 
appear to preclude the possibility of Charon formation from a graze-and-merge 
collision \citep[e.g.,][and references therein]{canup2005,canup2011,mckinnon2017,
bierson2018,stern2018}, it is useful to explore the evolution of a massive, 
circum-Pluto disk of debris to test alternate formation models 
\citep[e.g.,][]{desch2015,desch2017}.  Our goal is to learn whether the 
physical conditions required for the growth of a Charon-mass satellite are 
consistent with the structures generated by SPH calculations of giant impacts 
\citep{canup2005,canup2011}.

Tracking the growth of Charon within a circum-Pluto disk also places limits 
on plausible hit-and-run models. As we show below, Charon analogs reach their
final mass on short times scales, $\sim$ weeks.  At this point, the Charon 
analog lies within an expanding cloud of debris on an orbit similar to the 
published end states of hit-and-run impacts. The evolution of this debris as 
a function of initial conditions helps us to understand 
(i) the amount of material Pluto and Charon might accrete well after the 
impact and (ii) outcomes that are more (or less) favorable to the formation 
of the circumbinary satellites as the \pc\ binary expands from tidal 
interactions \citep[see also][]{walsh2015}.

Following a description of numerical methods (\S\ref{sec: sims}), we summarize the 
growth of Charon analogs from a disk of massive planetesimals (\S\ref{sec: res1})
and the dynamical evolution of massless tracers responding to the growth of a 
massive satellite (\S\ref{sec: res2}). After discussing how the results impact our 
understanding of the formation and evolution of the \pc\ system (\S\ref{sec: disc}),
we conclude with a brief summary.

\vskip 6ex
\section{NUMERICAL METHODS}
\label{sec: sims}

To track the evolution of circumplanetary debris in the Pluto system, we run a
series of numerical simulations with \orch, a parallel \verb!C++/MPI! hybrid 
coagulation + \nbody\ code that follows the accretion, fragmentation, and orbital 
evolution of solid particles ranging in size from a few microns to thousands of 
km \citep{kenyon2002,bk2006,kb2008,bk2011a,bk2013,kb2016a,knb2016}.  The ensemble 
of codes within \orch\ includes a multi-annulus coagulation code, an \nbody\ code, 
and radial diffusion codes for solids and gas.  Several algorithms link the codes 
together, enabling each component to react to the evolution of other components.

Here, we use the \nbody\ code to track the time evolution of the orbits of massive 
particles and massless tracers orbiting Pluto.  In the calculations, the tracers
serve as analogs of small particles of debris with negligible mass compared to 
the mass of Charon, \mc\ = $1.586 \times 10^{24}$~g.  Pluto has initial mass 
\mp\ = $1.303 \times 10^{25}$~g and radius \rp\ = 1183~km \citep{stern2015,nimmo2017}.  
Calculations begin with an ensemble of massive particles with total mass \M0\ and 
$N_t$ = 14000 tracers in orbit around Pluto.  Each massive particle has initial 
mass \m0, radius \r0, mass density $\rho_0$ = 1.75~\gcmc, and a semimajor axis, 
$\a0 \approx$ = 3--11~\rp.  Roughly half of the tracers are randomly placed in 
semimajor axis among the massive particles.  The rest have a random $a$ = 10--55~\rp. 

For the adopted physical properties of Pluto and the orbiting solids, the 
fluid Roche limit lies at $a_R \approx$ 2.5~\rp\ \citep[e.g.,][]{aggarwal1974,
weiden1984a,canup1995,hyodo2014}. Material at the Roche limit has an orbital 
period $P \approx$ 10~hr.  Solid particles with $a \lesssim a_R$ relax to a 
non-spherical equilibrium shape that depends on $a$ and the material strength 
\citep[e.g.,][]{holsapple2006,holsapple2008}. Although collision rates depend on
geometric cross-sections, we ignore modest differences in rates for ellipsoidal
particles in this pilot study. For the 150--250~km solid particles considered 
here, self-gravity sets the strength against tidal stresses. These particles are
probably stable against tides at $a \gtrsim 0.7 a_R \gtrsim$ 2~\rp. Although 
orbital motion can place solids inside this limit, 150--250~km solids with large 
gravity are more likely to experience surface and perhaps internal fractures than 
tidal shredding \citep[e.g.,][]{holsapple2006,sharma2006,holsapple2008,sharma2009,
quill2016}. During the much longer period of time spent outside the Roche limit, 
they probably recover their original structure. Thus, we ignore the possibility
of tidal shredding when solids lie inside the Roche limit.

Collision outcomes also depend on tidal stresses \citep[e.g.,][]{canup1995,
ohtsuki2013,hyodo2014}. Two colliding particles with masses $m_1$ and 
$m_2 = q m_1$ cannot merge if the sum of their radii is larger than their 
mutual Hill sphere,
\begin{equation}
\rhill = \left ( \frac{m_1 ~ + ~ m_2}{3 ~ \mp} \right )^{1/3} ~ \bar{a} 
     ~ = ~ \frac{r_1}{\rp} \left ( \frac{1 + q}{3} \right )^{1/3} ~ \bar{a} ~ ,
\label{eq: rhill}
\end{equation}
where $\bar{a} = 0.5 (a_1 + a_2)$.  Setting $\bar{a} = \alpha ~ \rp$ and 
$\tilde{r} = (r_1 + r_2) / \rhill \lesssim 1$, accretion is possible when
\citep{hyodo2014}:
\begin{equation}
\label{eq: r-acc}
\alpha \gtrsim \left ( \frac{3 \rho_P}{\rho_0} \right) ^{1/3} \frac{1 + q^{1/3}}{(1 + q)^{1/3}} ~ .
\end{equation}
With $\rho_P \approx \rho_0$, $\alpha \approx$ 2.3 for $q$ = 1;
$\alpha \approx$ 1.4 for $q$ = 0 \citep[see also][]{weiden1984a}. 
\citet{canup1995} derive a similar criterion. In our calculations, 
collisions between two \nbodies\ inside this limit are statistically 
unlikely. After the conclusion of each calculation, we identify 
\nbodies\ with positions inside 2.5~\rp\ and verify that they do not collide 
with other \nbodies\ while at a distance $r \lesssim$ 2.5~\rp. For the suite 
of simulations discussed here, no \nbody\ violates these constraints.  

Compared to the final states of SPH calculations for graze-and-merge collisions 
\citep[for example, Fig.~1 of][]{canup2005}, our initial conditions are somewhat
more radially extended.  Roughly 24~hr after the collision, the SPH calculations 
show solids extending from close to the surface of the central planet, 1--2~\rp, 
to $\sim$ 10--15~\rp; 70\% of this material initially lies outside the Roche limit.
Because the central planet has a short rotational period, $\sim$ 2.5~hr, solids 
that orbit synchronously with the planet lie well inside the Roche limit at 
$a_c \approx$ 1.05~\rp.  Over time, material orbiting the central planet eventually 
evolves into a disk \citep{canup2005}.  To avoid calculating the more complicated 
evolution of solids inside the Roche limit \citep[e.g.,][]{canup1995,ohtsuki2013,
hyodo2014}, the disks considered here have an inner radius $a_{in} \approx$ 3~\rp, 
an outer radius $a_{out} \approx$ 11~\rp, and a surface density distribution 
$\Sigma \propto a^{-n}$ with $n \approx$ 1.7--1.8.  Although \citet{canup2005} 
does not quote $\Sigma(a)$ for circum-Pluto disks, \citet{ida1997} consider 
similarly shallow surface density distributions in \nbody\ calculations of the 
formation of the Moon from an extended disk surrounding the Earth. Consistent with 
energy equipartition through gravitational scattering as circum-Pluto material 
evolves into a disk \citep{brahic1976,stewart1988, oht2002}, massive particles 
and tracers have the same initial eccentricity \e0\ and inclination \e0/2 
\citep[see also][]{ida1997}.

Throughout a calculation, the adaptive integrator adjusts time steps to resolve 
collisions (i) among all massive particles (including Pluto) and (ii) between 
tracers and other massive particles.  Although tracers may merge with massive 
particles, the massive particles do not respond to the tracers.  When a tracer 
or massive particle has $a \gtrsim \amax$ or $e > 1$, the code flags it as 
ejected and freezes its position and velocity for the remainder of the 
calculation. In these calculations, $\amax$ = 5400~\rp, slightly smaller
than the Hill radius of the \pc\ binary planet, $R_{H,PC} \approx$ 6750~\rp.

To avoid starting with a gravitationally unstable ring of massive particles, 
calculations begin with a sufficiently large \e0. Defining the initial velocity
dispersion \v0, angular velocity $\Omega_0$, and surface density $\Sigma_0$, a 
ring is unstable when $\v0 \Omega_0 < \pi G \Sigma_0$ \citep[e.g.,][]{chiang2010}.
To cast this condition in terms of \mp, \e0, and \M0, we set 
(i) $ \M0\ = \pi \Sigma_0 a_0^2$, where $a_0$ is a characteristic semimajor axis
for the swarm, and
(ii) $\v0\ = e v_K$, where $v_K$ is the local orbital velocity. The stability 
condition is then $\e0 \gtrsim M_0 / \mp$. Requiring that the mass reservoir
orbiting Pluto be sufficient to produce a Charon ($M_0 \gtrsim 0.1 \mp$), 
$\e0 \gtrsim$ 0.1.

In this set of calculations, we ignore fragmentation when two large objects 
physically collide and merge. For a collision involving two icy objects with
$r$ = 100~km on orbits with $a \sim$ 5--6~\rp\ and $e \sim 0.2$, the typical 
mass lost to debris is less than 1\% \citep[][and references therein]{kb2014b,
kb2015b,kb2017a}. To simplify this initial set of calculations, we ignore the 
impact of this debris on the evolution of massive and massless particles in 
the \nbody\ code. After each calculation, we verify that collisions 
between \nbodies\ are at velocities insufficient to eject more than 1\% of 
the combined mass of colliding objects in debris that escapes the merged 
object.  This condition is more stringent than the velocity limits required
for mergers inside the Roche limit \citep[e.g.,][]{hyodo2014}.

To survey outcomes as a function of initial conditions, we consider a range of 
\m0, \e0, and \M0. The range of total mass \M0\ -- 0.85~\mc\ to 1.75~\mc\ -- 
brackets the current mass of Charon.  We choose \e0\ = 0.1, 0.2, 0.3 or 0.4 
to cover many outcomes from detailed SPH calculations of giant impacts 
\citep{canup2005,canup2011}.  Although rings of particles with $\e0\ \approx$ 
0.1 are formally unstable for \M0\ = 1.75~\mc, these calculations allow us to 
examine trends in outcomes as a function of \e0\ in more detail.  The three 
choices for \r0\ -- 145~km, 185~km, and 230~km -- yield a manageable number of 
initial massive particles and allow us to learn whether outcomes depend on \r0.  
With 56 cores, each calculation requires a few hours to several days of computer 
time. We perform 12--15 calculations with each setup to test the sensitivity 
of outcomes to shot noise.

Comparing the final states of massless tracers with the \pc\ circumbinary 
satellites requires (i) knowledge of the physical properties of the satellites and 
(ii) an understanding of the processes required to convert debris left over from
the formation of Charon into a few small satellites. From observations with HST
\citep{brozovic2015,showalter2015}, the satellites have nearly circular orbits in 
the orbital plane of the \pc\ binary with semimajor axes 
$a_S \approx$ 36.06~\rp\ (Styx), $a_N \approx$ 41.16~\rp\ (Nix),
$a_K \approx$ 48.84~\rp\ (Kerberos), and $a_H \approx$ 54.72~\rp\ (Hydra).
Direct \nbody\ calculations place robust constraints on the masses of Nix
and Hydra \citep{kb2019b}: $m_N \lesssim 4.7 \times 10^{19}$~g and
$m_H \lesssim 5.5 \times 10^{19}$~g. Although the masses of Styx and Kerberos 
have larger uncertainties, the \nbody\ calculations suggest total masses of the
circumbinary satellite system $M_{SNKH} \lesssim 10^{20}$~g \citep{kb2019b}. Together 
with size measurements from \nh\ \citep{weaver2016}, these results suggest the 
circumbinary satellites have mass densities smaller than Pluto and Charon, 
$\rho_{SNKH} \lesssim$ 1.5~\gcmc.

\vskip 6ex
\section{RESULTS: EVOLUTION OF MASSIVE PARTICLES }
\label{sec: res1}

In all calculations, the sequence of merger events follows one of several well-defined
paths. Often, several early collisions generate 3--4 objects with masses equal to 
2~\m0.  These larger objects then accrete most of the other smaller planetesimals.
Sometimes, binary collisions among the $N$ initial objects generate $\sim N/2$ 
larger objects; mergers among this swarm eventually produce several large object(s) 
orbiting Pluto.  Rarely, one object rapidly merges with two other objects in the 
swarm and then accumulates material from the rest of the swarm. As the largest 
objects grow in mass, they often eject several small planetesimals from the system.
Once only 3--4 \nbodies\ remain in any of these scenarios, they either merge to 
form a single massive object, collide with Pluto, or are ejected from the system. 

\begin{figure}
\begin{center}
\includegraphics[width=5.0in]{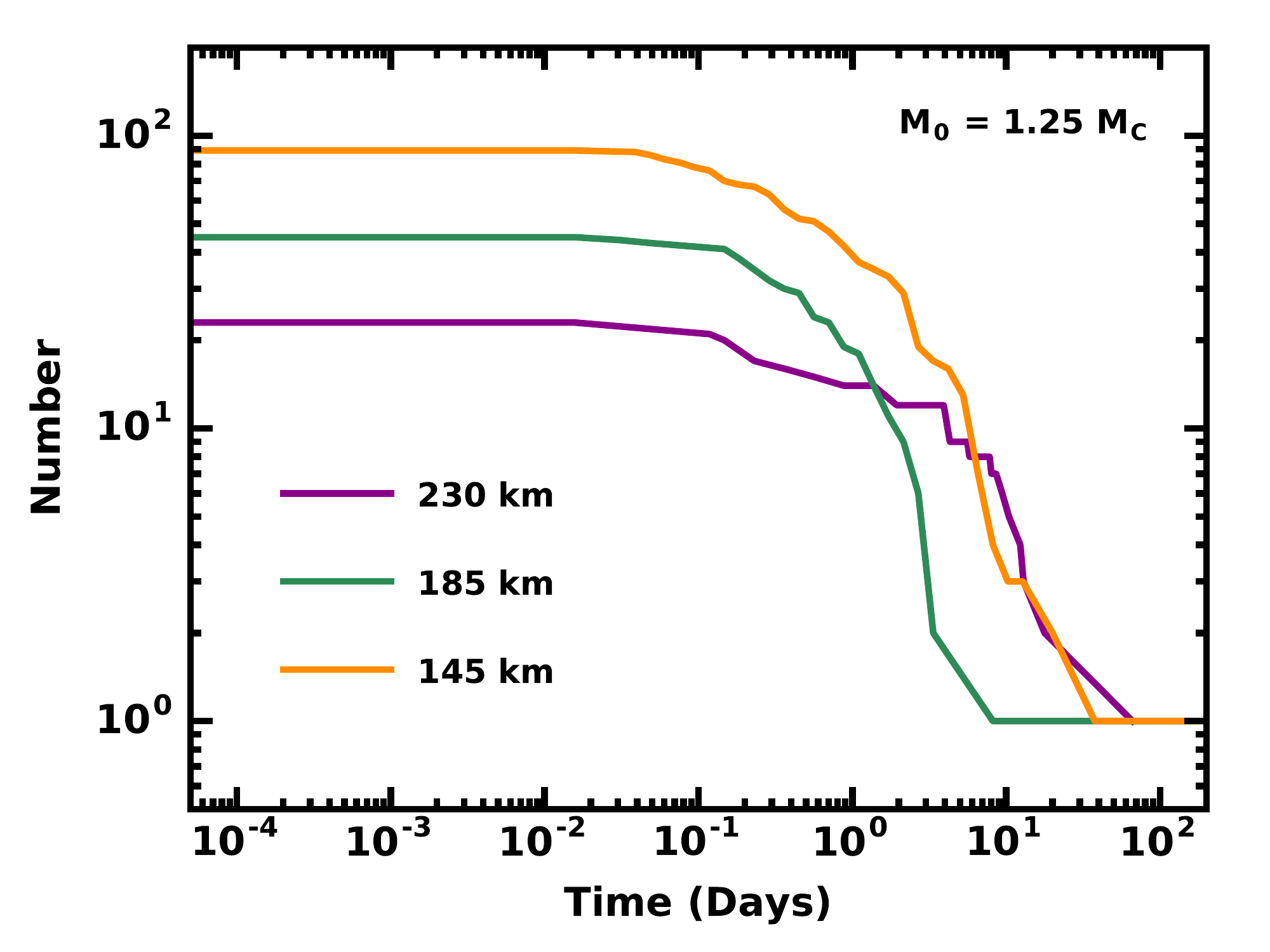}
\vskip -2ex
\caption{\label{fig: number}
Time evolution of the number of \nbodies\ in the swarm for calculations with 
\M0\ = 1.25~\mc\ and \r0\ = 145, 185, or 230~km as indicated in the legend.
Within a few hours, several \nbodies\ merge to form larger objects. On time
scales of 1--2~days (10--50~days), collisions reduce the number of \nbodies\ in
half (to 1--2). 
}
\end{center}
\end{figure}

As the largest \nbodies\ merge and grow, the dynamical evolution of the full ensemble
is rather simple. Within a few orbits, short-range scattering events raise (lower) 
the typical $e$ ($\imath$) of most \nbodies. After several mergers, dynamical friction 
tends to reduce the $e$ of the more massive \nbodies\ and raise the $e$ of the less 
massive \nbodies. Additional mergers accelerate this process, until several 
\nbodies\ have $e \approx$ 1. During the next 1--2 orbits, these \nbodies\ either
merge with another \nbody\ or are ejected. Once a single \nbody\ has more than 90\%
of the mass outside Pluto, it rapidly clears its orbit of any remaining \nbodies.
Across all of the calculations, Pluto and the most massive \nbodies\ eject up to nine 
low mass particles from the system; 3--4 ejections is typical.

Fig.~\ref{fig: number} illustrates the time evolution of the number of \nbodies\ in
typical calculations. After the first merger of two \nbodies, it takes only a day or two
for collisions to reduce  the number of \nbodies\ by half. After 10--50 days, only one 
(sometimes two) large object(s) orbit(s) Pluto. Among all of the calculations, the time 
scale for the number of \nbodies\ to fall to 0, 1, or 2 ranges from a few days to $\sim$ 
50--100 days. 

The outcomes of the evolution are fairly insensitive to the initial conditions. 
Although shot noise is important in every calculation, the time scale to reduce 
the number of \nbodies\ from $N$ to 1--2 scales inversely with \M0: more massive
collections of \nbodies\ evolve more rapidly than less massive swarms. However,
systems with \e0\ = 0.1 evolve as fast as systems with \e0\ = 0.4. Calculations 
with smaller planetesimals initially evolve somewhat more rapidly than those with 
larger planetesimals, but this difference is lost after several mergers generate
an ensemble of objects with masses 2-3 times larger than the initial mass.

\begin{figure}
\begin{center}
\includegraphics[width=5.0in]{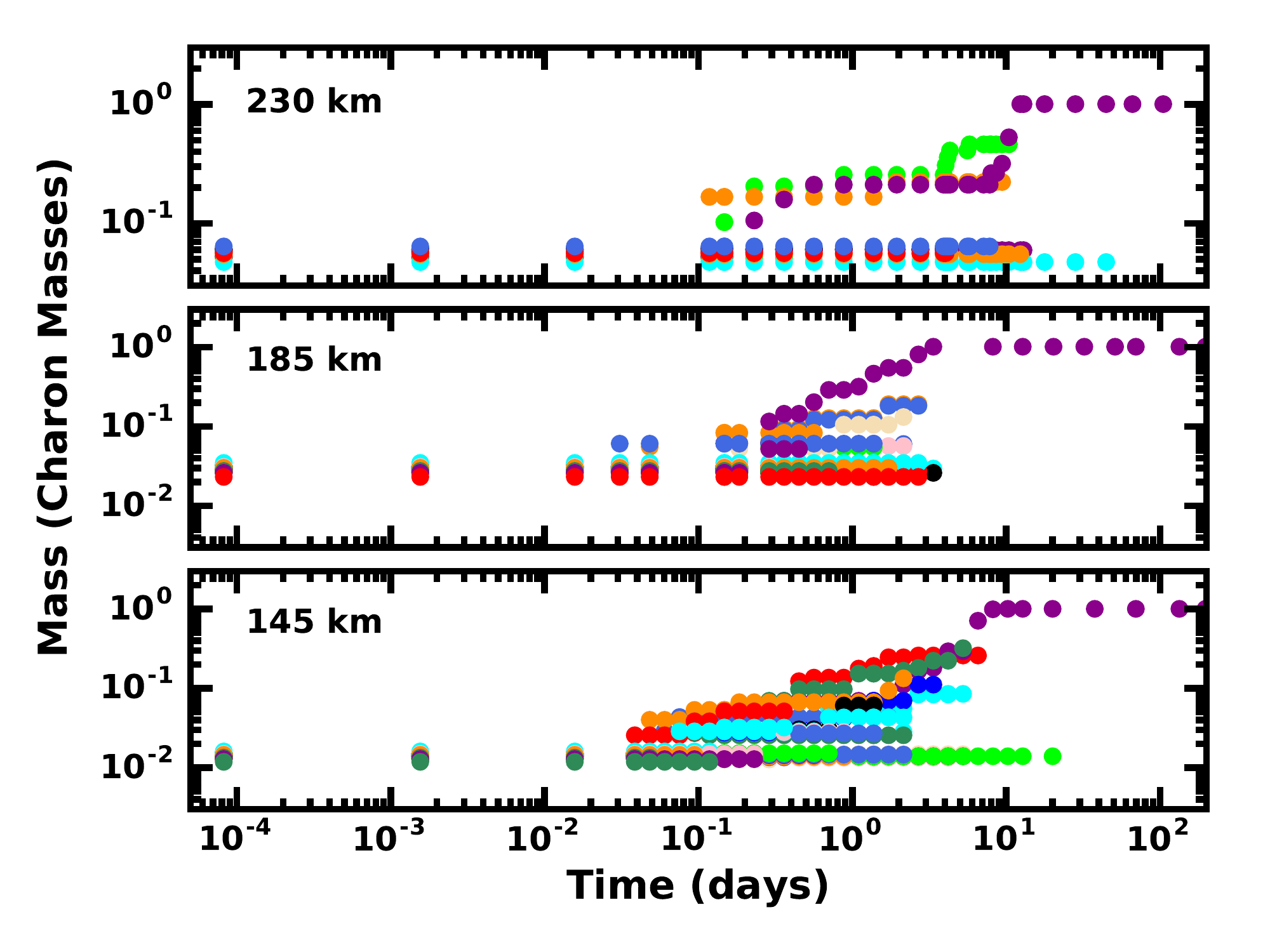}
\vskip -2ex
\caption{\label{fig: mass}
As in Fig.~\ref{fig: number} for the masses of surviving \nbodies. Each 
\nbody\ retains a distinctive color throughout the time frame shown in the
figure. Within 3--10 days, collisions among the \nbodies\ generate a single
object with a mass comparable to the mass of Charon.
}
\end{center}
\end{figure}

Fig.~\ref{fig: mass} shows the time evolution of the mass of each \nbody\ for the set of
calculations in Fig.~\ref{fig: number}. Initially, all of the \nbodies\ have the same 
mass. Within a few hours, several have merged with other \nbodies\ and doubled, tripled, 
or quadrupled in mass. In some cases, one \nbody\ gains mass more rapidly than the others
and remains the most massive object until the end of the calculation (Fig.~\ref{fig: mass},
middle panel). In others, several \nbodies\ have comparable masses throughout the evolution,
until one emerges as most massive (Fig.~\ref{fig: mass}, upper and lower panels). However,
the evolution proceeds, after 10--50 days only 0, 1, or 2 \nbodies\ orbit(s) Pluto.

Sometimes, 1--2 objects escape collisions for 10--100 days (Fig.~\ref{fig: mass}, green
object in the lower panel and the cyan object in the upper panel). Throughout their 
evolution, larger objects tend to stir their eccentricities to larger and larger values. 
Usually, the largest object in the swarm accretes this wayward planetesimal.  Sometimes, 
stirring leads to an ejection or to accretion by Pluto. 

\begin{figure}
\begin{center}
\includegraphics[width=5.0in]{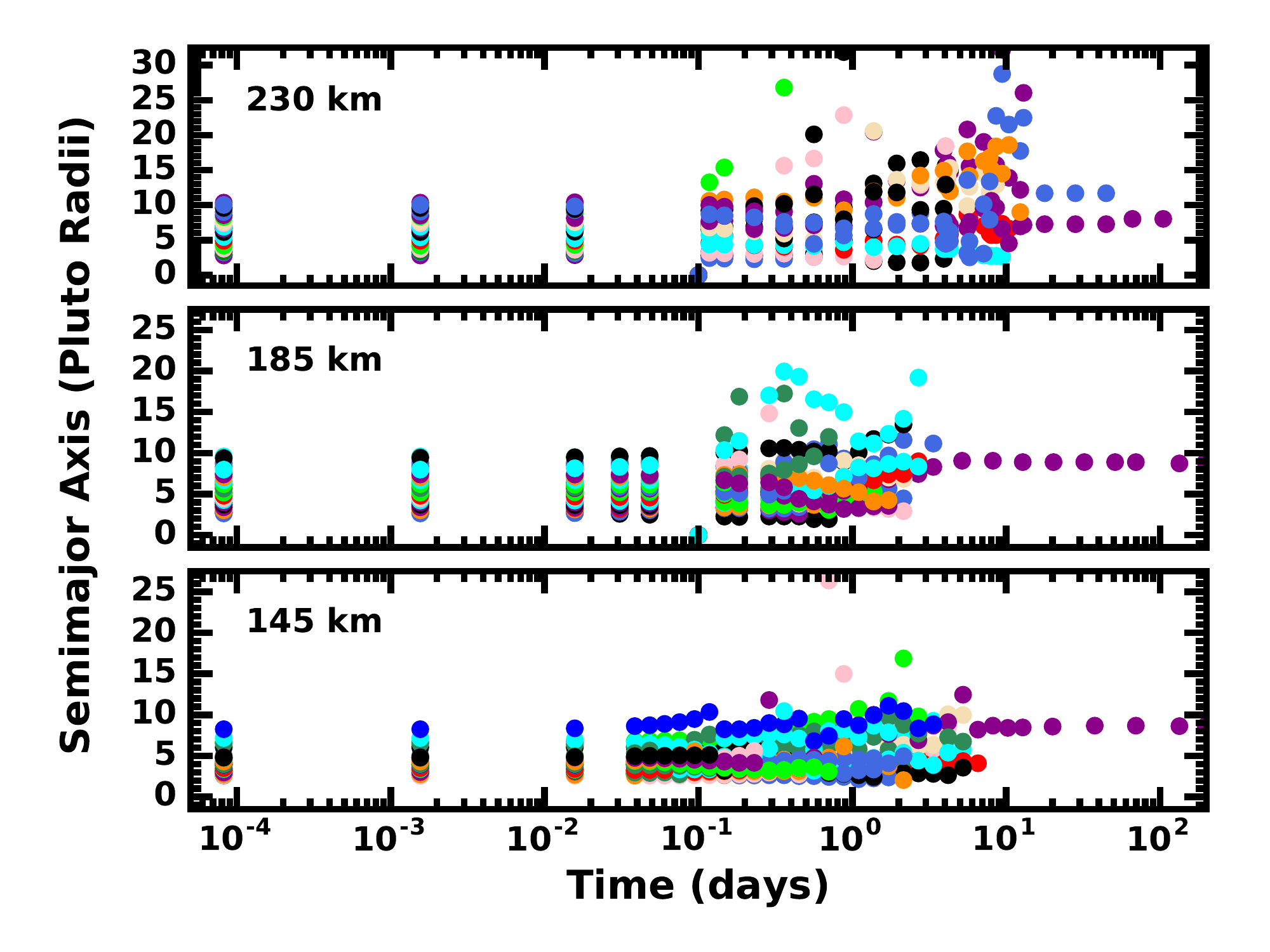}
\vskip -2ex
\caption{\label{fig: semi}
As in Fig.~\ref{fig: number} for the semimajor axis. Once a single massive
\nbody\ remains, it has a semimajor axis of 5--10~\rp\ and an orbital 
period of 1.25--3.5~days.
}
\end{center}
\end{figure}

Despite the chaotic evolution in mass, the most massive \nbody\ has a final orbit with
$a \approx$ 5--10~\rp\ (Fig.~\ref{fig: semi}). During the chaotic growth phase when
the number of \nbodies\ declines dramatically, the semimajor axes (eccentricity) of each 
\nbody\ vary from $\sim$ 3~\rp\ to 15--25~\rp\ ($\lesssim$ 0.1 to $\gtrsim$ 0.5). 
Sometimes, interactions between the last 2--3 \nbodies\ leads to impacts on Pluto, ejections
from the system, and no surviving companions to Pluto. Usually, however, a single massive
companion remains in orbit at 5--10~\rp\ from Pluto.

Fig.~\ref{fig: semi} also illustrates the different types of semimajor axis evolution for
the swarm of \nbodies. In the top panel, the larger initial masses of individual objects
stir up the orbits of their nearest neighbors, expanding the extent of the swarm in $a$.
Although the orbits then begin to cross, few \nbodies\ move from the inner edge of the 
swarm to the outer edge of the swarm over the course of the evolution. For the calculation
shown lower panel, there is less stirring among lower mass objects; orbit crossing is more 
common. The semimajor axis evolution is more chaotic, with several small objects moving 
from small $a$ to large $a$ (and sometimes back to small $a$) as the largest objects grow. 

In all of the calculations, several objects reach $a \gtrsim$ 20--25~\rp\ with large $e$
(e.g., the green object in the top panel of Fig.~\ref{fig: semi} at 0.3--0.4 days). These
objects are rarely accreted by Pluto or another object in the swarm. Usually, they are 
ejected from the system.

Statistics for the full ensemble of calculations reveals several characteristic outcomes
(Table~\ref{tab: masses}). Throughout the evolution, Pluto usually suffers one or more
large impacts from the orbiting swarm. The typical gain in Pluto's mass ranges from 
$\delta m_1 \approx 0.02 ~ \mp$ when $M_0 = 0.85 ~ \mc$ to 
$\delta m_1 \approx 0.05 ~ \mp$ when $M_0 = 1.75 ~ \mc$. The dispersion in this mass gain 
is comparable to the mass gain. In some calculations, Pluto accretes material roughly 
equivalent to the mass of Charon. 

The final mass \mtwo\ of the largest \nbody\ orbiting Pluto correlates with the initial mass 
in solids. When $M_0 = 0.85 ~ \mc$, \mtwo\ is roughly half the mass of Charon. As the initial
mass in solids grows, \mtwo\ also grows, reaching 5\% larger than the mass of Charon when
$\M0 = 1.75 ~ \mc$. The ratio $\mtwo\ / \M0$ also correlates with \M0, declining from 0.65 
when \M0\ is small to 0.57 when \M0\ is large. Calculations with small \M0\ are more 
efficient in concentrating the initial mass into a single large \nbody\ orbiting Pluto.

\begin{deluxetable}{lccccc}
\tablecolumns{6}
\tablewidth{0pc}
\tabletypesize{\footnotesize}
\tablenum{1}
\tablecaption{Results for \nbody\ calculations\tablenotemark{a}}
\tablehead{
  \colhead{$M_0 / m_C$} &
  \colhead{$r_0$ (km)} &
  \colhead{$\mu_1$} &
  \colhead{$\mu_2$} &
  \colhead{$a_2$ (\rp)} &
  \colhead{$e_2$} 
}
\label{tab: masses}
\startdata
0.85 & 230 & 1.02 $\pm$ 0.02 & 0.61 $\pm$ 0.07 & 6.25 $\pm$ 1.75 & 0.26 $\pm$ 0.12 \\
0.85 & 185 & 1.02 $\pm$ 0.02 & 0.54 $\pm$ 0.07 & 4.90 $\pm$ 0.97 & 0.20 $\pm$ 0.06 \\
0.85 & 145 & 1.02 $\pm$ 0.02 & 0.48 $\pm$ 0.10 & 5.02 $\pm$ 1.46 & 0.19 $\pm$ 0.07 \\
1.25 & 230 & 1.02 $\pm$ 0.01 & 0.79 $\pm$ 0.11 & 5.39 $\pm$ 1.40 & 0.20 $\pm$ 0.08 \\
1.25 & 185 & 1.03 $\pm$ 0.02 & 0.75 $\pm$ 0.10 & 6.74 $\pm$ 2.64 & 0.20 $\pm$ 0.06 \\
1.25 & 145 & 1.02 $\pm$ 0.02 & 0.72 $\pm$ 0.14 & 5.39 $\pm$ 1.62 & 0.21 $\pm$ 0.08 \\
1.75 & 230 & 1.04 $\pm$ 0.03 & 1.04 $\pm$ 0.31 & 6.68 $\pm$ 1.29 & 0.28 $\pm$ 0.10 \\
1.75 & 185 & 1.05 $\pm$ 0.04 & 1.02 $\pm$ 0.20 & 6.52 $\pm$ 1.62 & 0.26 $\pm$ 0.10 \\
1.75 & 145 & 1.04 $\pm$ 0.02 & 0.98 $\pm$ 0.14 & 6.04 $\pm$ 1.83 & 0.23 $\pm$ 0.06 \\
\enddata
\tablenotetext{a}{The columns list the ratio of the initial mass of the swarm relative
to the mass of Charon, $\M0/\mc$, the initial radius of solids orbiting Pluto \r0, the
ratio of the final to initial mass for Pluto ($\mu_1 = m_1 / \mp$), the ratio of the 
mass of the Charon analog to Charon's mass ($\mu_2 = m_2 / \mc$), and the average 
semimajor axis and eccentricity of the Charon analog and their dispersions.
}
\end{deluxetable}

The final mass also correlates with the initial sizes of the solids. Ensembles of small
solids are less efficient at producing a massive \nbody\ than ensembles of large solids
(Table~\ref{tab: masses}). When \r0\ is smaller, it is easier for the largest \nbody\ in
the swarm to scatter the smallest \nbody\ away from the rest of the swarm. Repeated 
dynamical interactions between the largest and smallest \nbody\ often result in the 
ejection of the smallest \nbody. When \r0\ is larger, there is less of a contrast between
the masses of the smallest and largest \nbodies, limiting the number of ejections. Fewer
dynamical ejections allow the largest \nbody\ to reach a larger \mtwo.

\begin{figure}
\begin{center}
\includegraphics[width=5.0in]{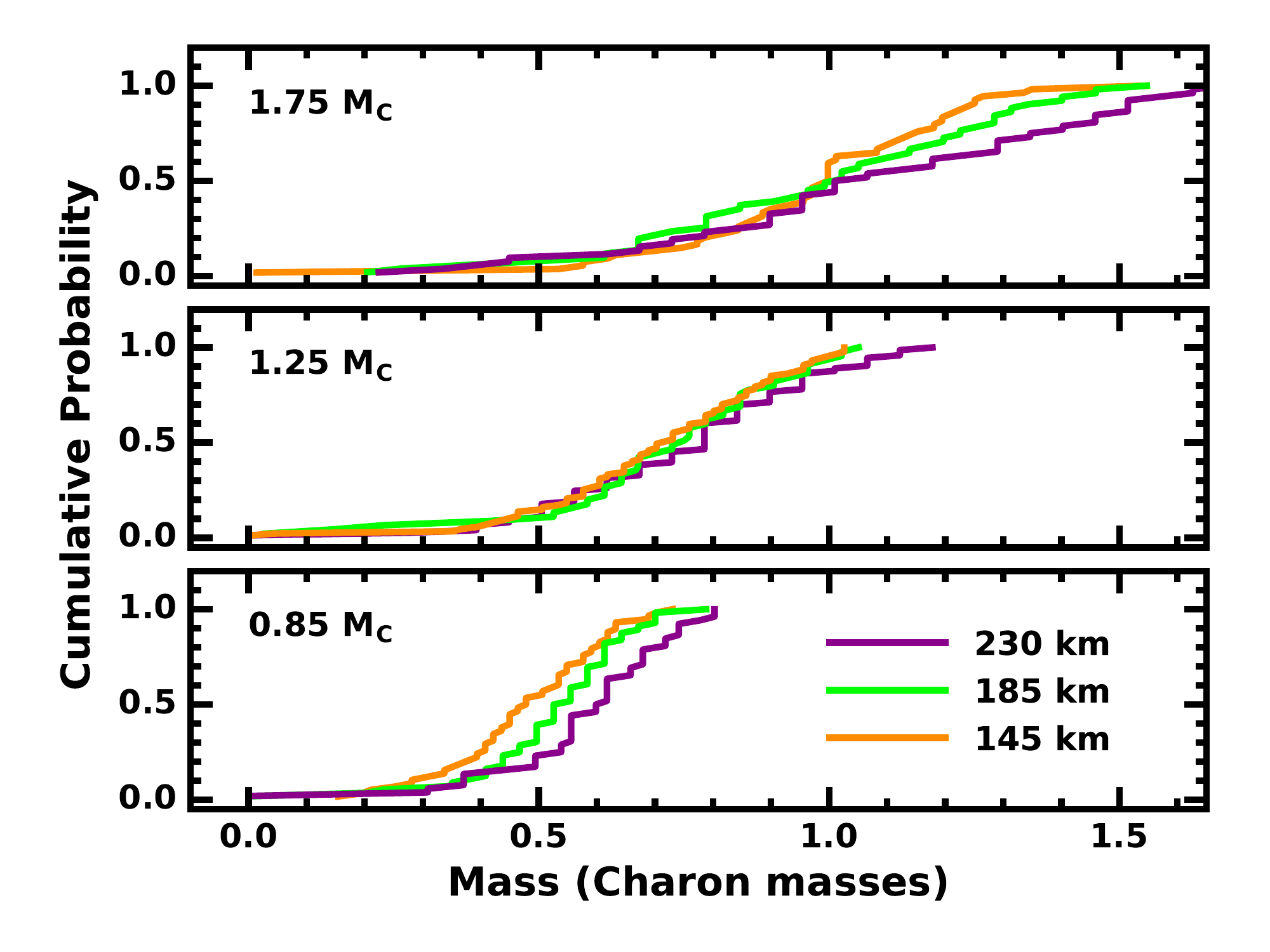}
\vskip -2ex
\caption{\label{fig: prob}
Cumulative probability for the final mass of the largest \nbody\ 
($p (\mtwo\ < m)$) for calculations with three different \M0\ and three 
different \r0\ as indicated in the legend of each panel. Roughly half 
of the calculations yield $\mtwo\ \lesssim 0.55 ~ \mc$ ($\M0\ = 0.85 ~ \mc$), 
$\mtwo\ \lesssim 0.75 ~ \mc$ ($\M0\ = 1.25 ~ \mc$), or
$\mtwo\ \lesssim 1.05 ~ \mc$ ($\M0\ = 1.75 ~ \mc$). As summarized in the text,
the probability of achieving $\mtwo\ \approx \mc$ depends on the initial
radius of massive objects in the swarm.
}
\end{center}
\end{figure}

Comparisons among the mass distributions suggest correlations with \r0\ but not 
\e0\ (Fig.~\ref{fig: prob}). When \M0\ = 0.85~\mc\ (Fig.~\ref{fig: prob}, lower panel),
the mass distributions for \r0\ = 145~km (orange curve), \r0\ = 185~km (green curve), 
and \r0\ = 230~km (purple curve) clearly differ from one another. Using a K--S test
\citep{press1992}, the probability that these distributions are drawn from the same 
parent population is negligible, $\lesssim 10^{-5}$ to $10^{-4}$. However, mass 
distributions for calculations with \M0\ = 1.25~\mc\ are nearly identical, with
large K--S probabilities, 30\% to 90\%, of being drawn from the same parent population. 
At \M0\ = 1.75~\mc, the mass distributions for the three \r0\ have a 5\% to 10\%
probability of being drawn from the same parent population.

Despite the variation of \mtwo\ with initial conditions, there is remarkably little 
variation of the final semimajor axis \atwo\ or eccentricity \etwo. Typically,
$\atwo\ \approx$ 5--7~\rp\ and $\etwo \approx$ 0.2.  With a dispersion 
of 1.0--2.5~\rp, the overall range in \atwo\ is large, with a minimum of 
3~\rp\ and a maximum of 11.5~\rp.  In some calculations, the final orbit 
of Pluto's companion is nearly circular, with $\etwo \approx$ 0.05--0.10. In others, 
the final orbit is highly eccentric $\etwo \approx$ 0.4--0.5.

Analyses of the variation of \atwo\ or \etwo\ with initial conditions yields no strong correlations. 
There is a weak correlation between \etwo\ and \e0, where systems with smaller \e0\ produce a
massive satellite with smaller \etwo. However, the trend is significant at less than the 
1.5-$\sigma$ level.  The final eccentricity of a massive satellite is insensitive to \M0\ or
\r0; \etwo\ ranges from 0.05 to 0.40 with a typical \etwo\ $\approx$ 0.2. Among all of the massive 
satellites produced in the calculations, \atwo\ is remarkably uncorrelated with \e0, \r0, or \M0. 
For any combination of \e0, \r0, and \M0, \atwo\ ranges from 3~\rp\ to 10--11~\rp,
with a typical \atwo\ $\approx$ 5--6~\rp.

\vskip 6ex
\section{RESULTS: EVOLUTION OF MASSLESS TRACER PARTICLES}
\label{sec: res2}

The swarm of massless tracer particles evolves in step with the growth of the largest \nbodies.
At the start of each calculation, the tracers adjust their orbits to the gravitational field of
all of the massive objects. Within a few orbits, some tracers impact Pluto or one of the massive 
\nbodies. Strong gravitational interactions lead to the ejection of several others from the system. 
The rest of the tracers settle down into eccentric orbits around Pluto.

As each system evolves, the orbits of the tracers gradually expand away from the orbits of the 
\nbodies. Early on, many tracers collide with Pluto or another \nbody. As the evolution proceeds,
the small dynamical kicks each tracer receives from the $N$ remaining \nbodies\ slowly scatters
them to larger and larger semimajor axes. With no collisional damping, the orbital eccentricities 
of many tracers gradually grow with time. Eventually, these tracers are ejected. Other tracers are
able to maintain a fairly small $e$ and end up on wide orbits around the central \pc\ binary.

\begin{figure}
\begin{center}
\includegraphics[width=5.0in]{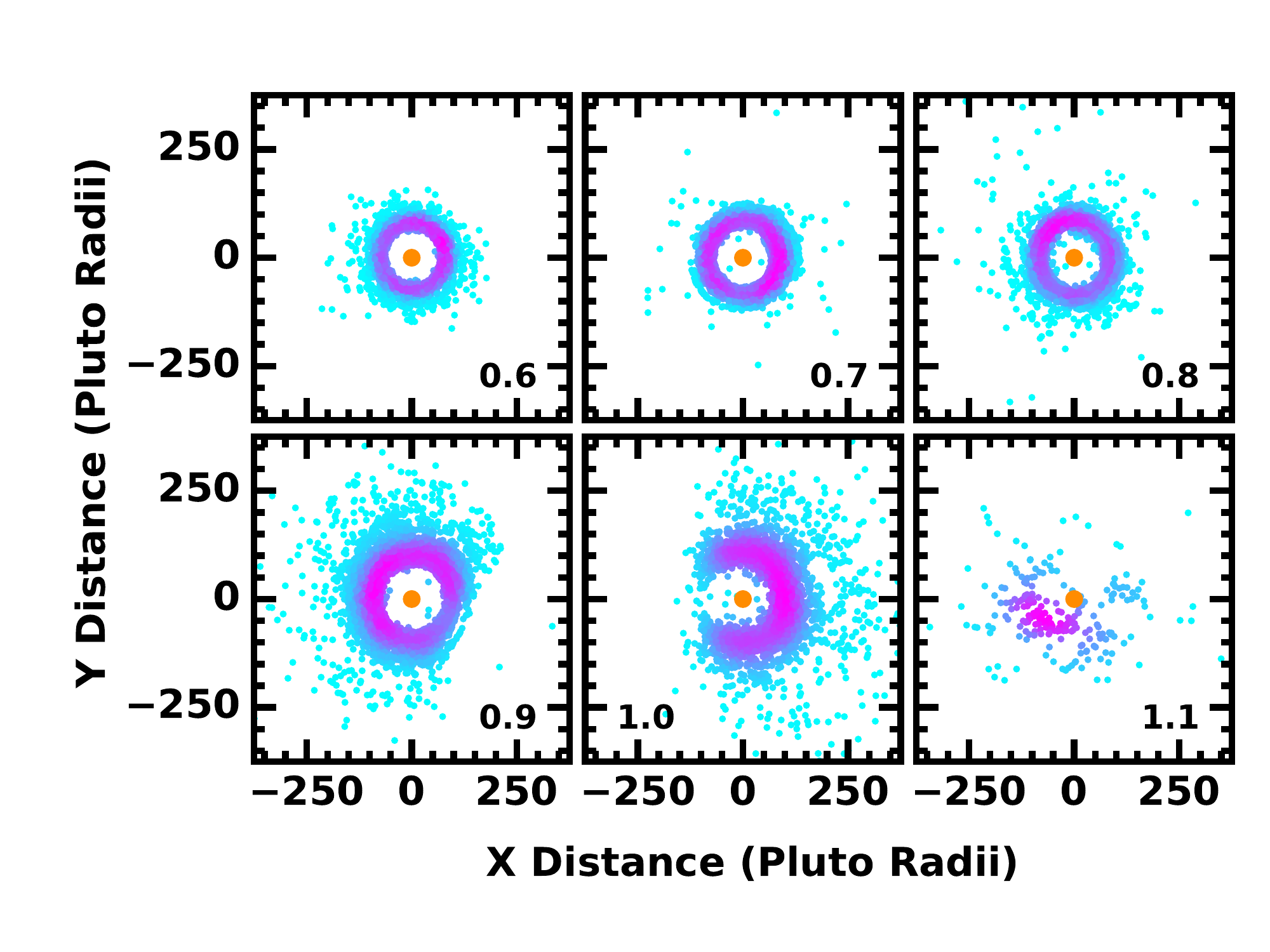}
\vskip -2ex
\caption{\label{fig: ring1}
Positions of massless tracers in the $x-y$ plane at 150--300~d. 
In each panel, the density of tracers in arbitrary units ranges f
rom low (cyan) to medium (purple) to high (magenta). The final mass 
of the Charon analog (in Charon masses) appears in the lower left or 
lower right corner of each panel. 
}
\end{center}
\end{figure}

Throughout the radial expansion and ejection of tracers, there is little excitation of tracer 
orbits in $z$. During the first few days of evolution, the vertical scale height of the tracers
is roughly constant. As they pass close to massive \nbodies, their eccentricities grow; however,
their inclinations are largely unchanged. In systems that develop structure in the $x-y$ plane 
(see below), there is little evidence for structure with $z$. Tracers with large $e$ are ejected
at inclinations fairly similar to their starting inclinations.

Fig.~\ref{fig: ring1} illustrates the positions of swarms of tracers in the $x-y$ plane at 
the end of several calculations. When the Charon analog has a small mass, $\mtwo\ \approx$ 
0.6--0.8~\mc\ (upper panel), the ensembles of tracers occupy fairly circular rings with radius 
of 100--175~\rp. Aside from a dense set of tracers in the well-defined rings, there is often 
a halo of tracers on orbits with larger semimajor axes, $a \approx$ 200--275~\rp. Although the 
rings of tracers look nearly circular, tracer orbits are eccentric, with typical $e \approx$ 
0.2--0.5.

In systems that produce a more massive Charon analog ($\mtwo\ \approx$ 0.9--1.1~\mc; 
Fig~\ref{fig: ring1}, lower panel), tracers suffer more energetic interactions with 
the \nbodies. Aside from having more frequent impacts with Pluto or other \nbodies, 
these systems eject more tracers.  In some cases, the few remaining tracers have a 
rather chaotic distribution in the $x-y$ plane (lower right panel).  In others, snapshots 
contain partial rings of tracers at distance of 100--275~\rp\ from the central binary 
(lower left and lower middle panels). Because the tracers from the empty portion of the ring 
have been ejected, the partial ring pattern appears to rotate as the evolution proceeds.

\begin{figure}
\begin{center}
\includegraphics[width=5.0in]{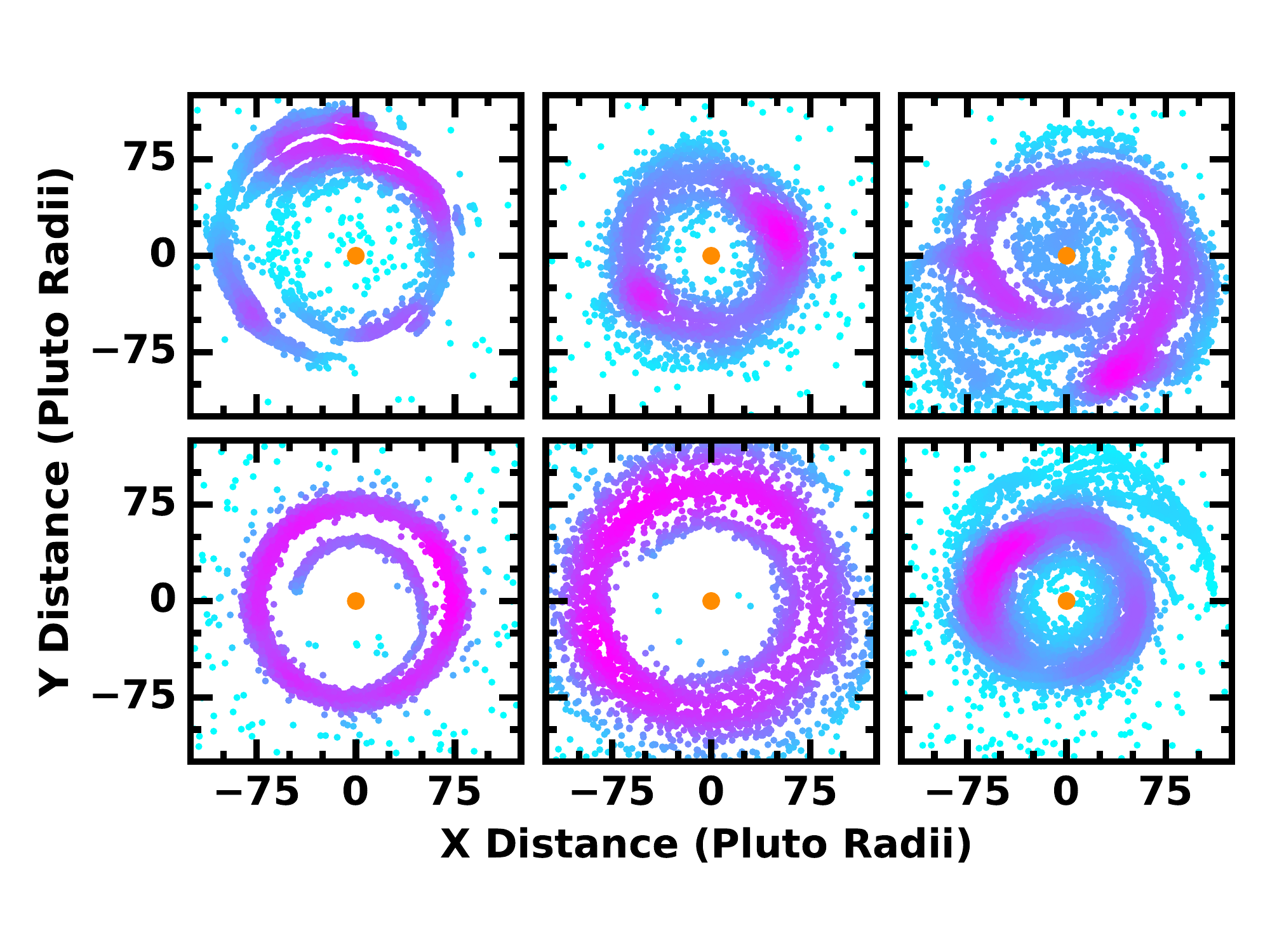}
\vskip -2ex
\caption{\label{fig: ring2}
As in Fig.~\ref{fig: ring1} for systems of tracers with large-scale spiral
structure at 150--300~d. Final masses for the Charon analog range from 
0.4~\mc\ to 0.8~\mc.
}
\end{center}
\end{figure}

Throughout many calculations, the distribution of tracers in the $x-y$ plane shows a clear 
spiral pattern (Fig.~\ref{fig: ring2}). These structures are often more compact than the rings 
displayed in Fig.~\ref{fig: ring1}, with typical radii of 50--100~\rp. Often, the tracers 
follow a clean one-armed spiral (lower left panel). Sometimes, the tracers lie in two distinct
concentrations, with a weak trailing spiral from each structure (upper middle panel). Other
ensembles of tracers show very tightly wound spirals (lower middle panel) or more open systems
resembling a spiral galaxy (upper right panel).

As with the ring systems in Fig.~\ref{fig: ring1}, systems with spiral structure tend to persist 
when the mass of the Charon analog is small, $\mtwo\ \lesssim$ 0.8~\mc.  When the final mass of 
the Charon analog is larger, short-lived spiral patterns are present as the mass of this analog 
grows with time. Once the mass exceeds $\sim$ 0.9~\mc, however, the Charon analog tends to eject 
small leftover massive particles out through the orbital plane. These massive particles take many
tracers with them, destroying the spiral pattern.

Despite the persistent ring or spiral structure among the tracers in many calculations, 
the typical positions of tracers are well outside the orbits of the current circumbinary 
satellites. In roughly half of the calculations, the semimajor axes for tracers at 
150--300~d are $a_t \sim$ 125--175~\rp, much larger than the semimajor axis of the 
outermost satellite Hydra, $a_H \approx$ 55~\rp. Only $\sim$ 10\% of systems retain 
tracers close to the orbit of Hydra. 

\begin{figure}
\begin{center}
\includegraphics[width=5.0in]{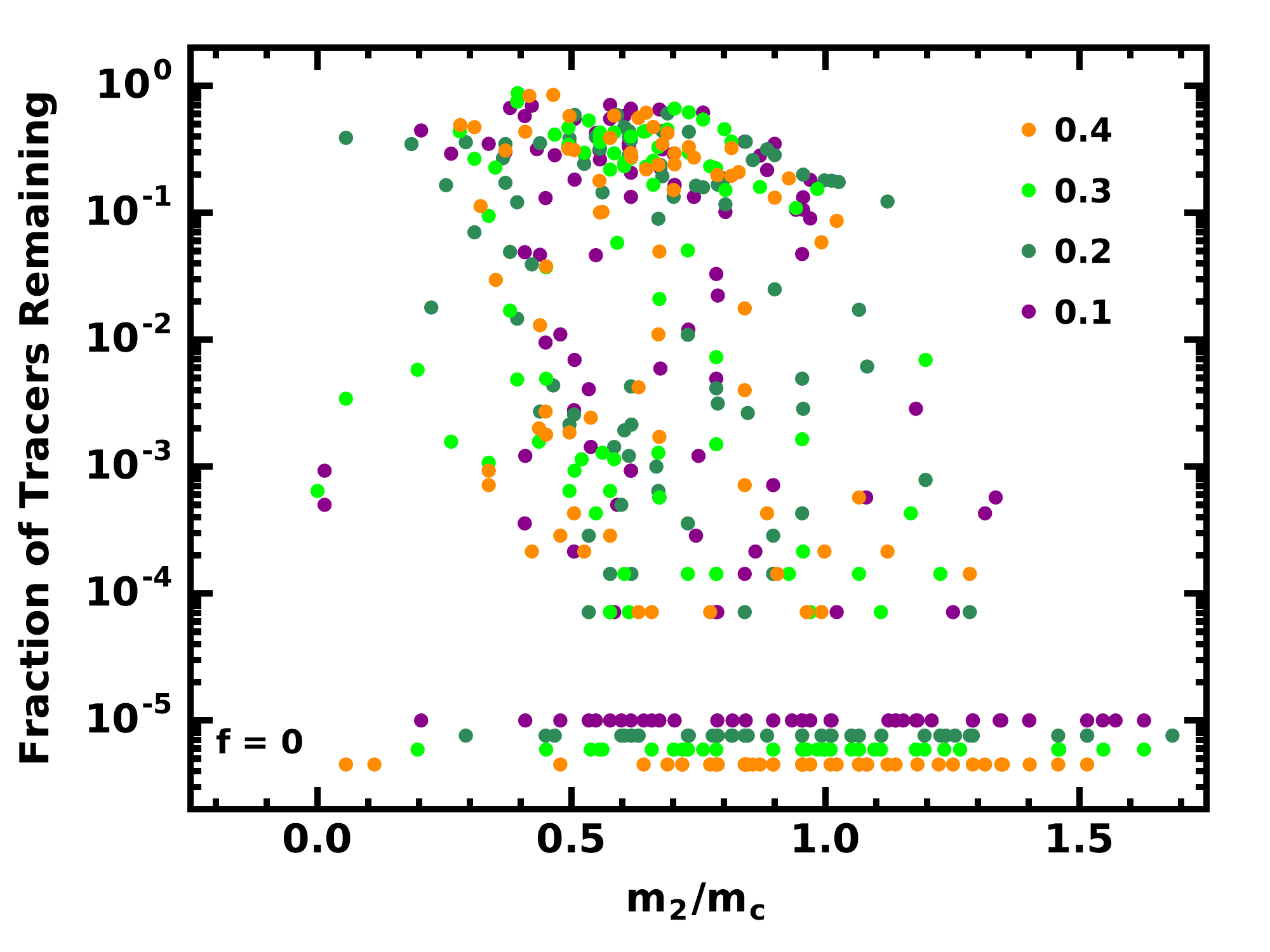}
\vskip -2ex
\caption{\label{fig: trace1}
Fraction of 14,000 tracers remaining after 150--300~d of dynamical evolution 
(when the most massive \nbody\ reaches its final mass). As noted in the legend,
symbols are color-coded according to $e_0$, the initial eccentricity of the 
\nbodies. The ability of a system to retain tracers is independent of $e_0$.
}
\end{center}
\end{figure}

Among all of the calculations, the fraction of tracers remaining in the system 
after 150--300~days, $f_t$, correlates with the final mass of the Charon analog 
(Fig.~\ref{fig: trace1}).  In the upper part of the plot, there is a clear 
progression from $f_t \approx$ 0.5--0.6 when $\mtwo\ \approx$ 0.4~\mc\ to 
$f_t \approx$ 0.1 when $\mtwo\ \approx$ \mc. For systems with fewer tracers,
$f_t \lesssim$ 0.1, there is a secondary trend where $f_t$ drops from $\sim$ 0.1 
at $\mtwo\ \approx$ 0.4~\mc\ to less than $10^{-3}$ at $m_2 \approx$ 0.6--0.7~\mc. 
From the color-coding of the points, these trends appear to be independent of $e_0$, 
the initial eccentricity of the \nbodies\ in the swarm.

Aside from the strong correlation of $f_t$ with $\mtwo$, the number of systems 
with few if any tracers is another striking feature of Fig.~\ref{fig: trace1}. 
Among the $\sim$ 475 calculations shown, $\sim$ 15\% have only 1--14 tracers 
left after $\sim$ 300 days of evolution. A much larger fraction, $\sim$ 33\%, 
have no tracers. These percentages correlate well with the final mass of the 
Charon analog. Of the $\sim$ 235 calculations where the final mass of the 
Charon analog is $\mtwo\ \lesssim$ 0.7~\mc, $\sim$ 15\% have no tracers 
after 150--300~d of evolution. In contrast, roughly 50\% of the $\sim$ 240 
calculations where $\mtwo\ \gtrsim$ 0.7~\mc\  have no tracers at the 
150--300~d mark (Table~\ref{tab: perc}, first two rows).

Survival of tracers depends less on the final semimajor axis or eccentricity 
of the Charon analog than on its mass (Table~\ref{tab: perc}, rows 3--6). 
Although there is a
slight tendency for systems with larger $a_2$ to retain more tracers, Charon 
analogs with $a_2 \lesssim$ 6~\rp\ and $a_2 \gtrsim$ 6~\rp\ are roughly 
equally likely to have $f_t$ = 0 or $f_t \gtrsim$ 0.1. The calculations 
display a similar trend with $e_2$: Charon analogs with $e_2 \lesssim$ 0.22 
are somewhat more likely to retain a larger fraction of tracers than those 
with $e_2 \gtrsim$ 0.22. However, the differences do not seem significant.

For any final orbit of the Charon analog, systems are much more likely to have
a large fraction of their initial complement of tracers or no tracers at all
than to have some intermediate fraction of tracers (Table~\ref{tab: perc}). 
Roughly equal numbers of systems (33\%) either have no tracers or most of their
tracers. Another 15\% have less than 0.1\% of their initial set of tracers; 
10\% (7\%) have 0.1\% to 1\% (1\% to 10\%). The relative lack of systems with
0.1\% to 10\% of their tracers provides a quantitative measure of the ability 
of dynamical ejections of massive \nbodies\ to remove tracers efficiently on 
short time scales. 

\begin{deluxetable}{lccccc}
\tablecolumns{6}
\tablewidth{0pc}
\tabletypesize{\footnotesize}
\tablenum{2}
\tablecaption{Tracer statistics at 150--300~d\tablenotemark{a}}
\tablehead{
  \colhead{Variable} &
  \colhead{$f(f_t = 0)$} &
  \colhead{$f(0 < f_t \le 10^{-3})$} &
  \colhead{$f(10^{-3} < f_t \le 10^{-2})$} &
  \colhead{$f(10^{-2} < f_t \le 10^{-1})$} &
  \colhead{$f(10^{-1} < f_t \le 1)$}
}
\label{tab: perc}
\startdata
$\mtwo\ \le$ 0.7~\mc\ & 0.14 & 0.15 & 0.15 & 0.09 & 0.48 \\
$\mtwo\ >$   0.7~\mc\ & 0.52 & 0.16 & 0.06 & 0.05 & 0.21 \\
$a_2 \le 6~\rp$ & 0.36 & 0.19 & 0.10 & 0.04 & 0.32 \\
$a_2 > 6~\rp$ &   0.31 & 0.12 & 0.10 & 0.10 & 0.38 \\
$e_2 \le 0.22$  &   0.32 & 0.19 & 0.07 & 0.05 & 0.37 \\
$e_2 > 0.22$    &   0.35 & 0.12 & 0.13 & 0.09 & 0.32 \\
\\
$e_0$ = 0.1 & 0.33 & 0.16 & 0.08 & 0.07 & 0.35 \\
$e_0$ = 0.2 & 0.32 & 0.12 & 0.12 & 0.07 & 0.36 \\
$e_0$ = 0.3 & 0.31 & 0.17 & 0.13 & 0.05 & 0.34 \\
$e_0$ = 0.4 & 0.37 & 0.16 & 0.07 & 0.07 & 0.33 \\
$r_0$ = 145~km & 0.28 & 0.16 & 0.08 & 0.06 & 0.42 \\
$r_0$ = 185~km & 0.32 & 0.17 & 0.11 & 0.08 & 0.33 \\
$r_0$ = 230~km & 0.41 & 0.15 & 0.11 & 0.06 & 0.28 \\
$M_0$ = 0.85~\mc\ & 0.19 & 0.14 & 0.13 & 0.07 & 0.47 \\
$M_0$ = 1.25~\mc\ & 0.16 & 0.16 & 0.09 & 0.07 & 0.52 \\
$M_0$ = 1.75~\mc\ & 0.66 & 0.17 & 0.08 & 0.06 & 0.04 \\
\enddata
\tablenotetext{a}{The columns list the physical variables and 
the fraction of calculations with the indicated ranges of $f_t$
for listed physical variable at an evolution time of 150--300~d.
}
\end{deluxetable}

To quantify correlations between $f_t$, the initial conditions for the 
planetesimals, and the final properties of the Charon analog in more 
detail, we rely on a parametric measure (Pearson's linear correlation
coefficient $r$) and two non-parametric measures 
\citep[the Spearman rank-order correlation coefficient, $r_S$, and Kendall's 
$\tau$;][]{press1992}. Table~\ref{tab: corr} lists the results.  Confirming 
the visual impression from Fig.~\ref{fig: trace1}, all three tests measure 
a strong correlation between $f_t$ and the final mass of the Charon analog $\mtwo$. 
Probabilities of no correlation range from $p_r \approx 4 \times 10^{-15}$ 
for the Pearson linear correlation coefficient to $p_\tau \approx 4 \times 10^{-30}$ 
for Kendall's $\tau$ to zero for the Spearman rank-order test.  

Although Kendall's $\tau$ suggests a 3$\sigma$ correlation between $f_t$ and 
$a_2$ (in the sense that Charon analogs with larger $a_2$ retain more tracers), 
the other two tests measure no significant correlation ($p_r$ = 0.51 and 
$p_S$ = 0.18; Table~\ref{tab: corr}).  The Spearman rank-order test indicates 
that the correlation between $f_t$ and $e_2$ barely has a 2$\sigma$ significance; 
the other tests measure no correlation ($p_r$ = 0.60 and $p_\tau$ = 0.90).  
Thus, the fraction of remaining tracers depends on the final mass of the 
Charon analog, but not its semimajor axis or eccentricity.

The correlation coefficients also generally confirm the visual impression 
that the number of tracers remaining after 150--300~d is independent of 
$e_0$. The Pearson ($p_r$ = 0.91) and Kendall ($p_\tau$ = 0.48) tests 
suggest very high probabilities that $e_0$ and $f_t$ are uncorrelated. 
Although the Spearman rank-order test indicates a very low probability 
that $e_0$ and $f_t$ are uncorrelated ($p_S \approx 10^{-11}$, with 
$r_S$ = $-0.31$), a plot of $f_t$ as a function of $e_0$ (not shown) 
looks completely uncorrelated. We suspect that the low probability from 
the Spearman rank-order test results from the marginally larger than 
average number of systems with $e_0$ = 0.4 that retain fewer than 10 
tracers after 150--300~d and the marginally smaller than average number 
of systems with $f_t \approx$ 0.001-0.1 (Table~\ref{tab: perc}).

In Fig.~\ref{fig: trace1}, there are several examples where the \nbodies\ have
ejected nearly all of the initial complement of tracers ($f_t \lesssim 10^{-2}$) 
despite the production of a very low mass Charon analog ($\mtwo\ \lesssim$ 0.1~\mc).
As the collisional and dynamical evolution proceeds, these systems initially 
produce a massive Charon analog ($\mtwo\ \gtrsim$ 0.75~\mc) with a small semimajor 
axis ($a_2 \lesssim$ 4~\rp) and a very eccentric orbit ($e_2 \gtrsim$ 0.4--0.5).  
Interactions with several smaller left over \nbodies\ generates an orbit with larger 
and larger $e_2$; eventually, the Charon analog collides and merges with Pluto.  
Although 1--2 low mass \nbodies\ may survive this evolution on high $e$ orbits 
at large $a$, most of the tracers are ejected. Given the small numbers, it is 
impossible to judge whether these systems are associated with any particular $e_0$.

\begin{deluxetable}{lcccccc}
\tablecolumns{7}
\tablewidth{0pc}
\tabletypesize{\footnotesize}
\tablenum{3}
\tablecaption{Correlations of Physical Variables with $f_t$ at 150--300~d\tablenotemark{a}}
\tablehead{
  \colhead{Variable} &
  \colhead{$r$} &
  \colhead{$r_S$} &
  \colhead{$\tau$} &
  \colhead{$p_r$} &
  \colhead{$p_S$} &
  \colhead{$p_\tau$} 
}
\label{tab: corr}
\startdata
$\mtwo$ & $-$0.35 & $-$0.59 & $-$0.35 & $4.0 \times 10^{-15}$ & 0.0                   & $ 4.1 \times 10^{-30}$ \\
$a_2$ & $-$0.03 & $+$0.06 & $+$0.10 & $5.1 \times 10^{-1}$  & $1.8 \times 10^{-1}$ & $ 9.3 \times 10^{-4}$ \\
$e_2$ & $-$0.02 & $-$0.09 & $+$0.00 & $6.0 \times 10^{-1}$  & $4.7 \times 10^{-2}$ & $ 9.0 \times 10^{-1}$ \\
$e_0$ & $+$0.01 & $-$0.31 & $-$0.02 & $9.1 \times 10^{-1}$  & $8.8 \times 10^{-12}$ & $ 4.8 \times 10^{-1}$ \\
$r_0$ & $-$0.14 & $-$0.63 & $-$0.11 & $1.7 \times 10^{-3}$  & 0.0                   & $ 2.7 \times 10^{-4}$ \\
$M_0$ & $-$0.31 & $-$0.91 & $-$0.35 & $2.1 \times 10^{-12}$ & 0.0                   & $ 5.9 \times 10^{-30}$ \\
\enddata
\tablenotetext{a}{The columns list the physical variables,
the correlation coefficients for the Pearson ($r$), 
Spearman rank-order ($r_S$) and Kendall ($\tau$) tests along 
with the probabilities for a lack of correlation between the
listed variable and $f_t$ at 150--300~d.
}
\end{deluxetable}

Although the ability of a system to retain tracers is independent 
of $e_0$, swarms that start the evolutionary sequence with larger 
\nbodies\ lose their tracers more rapidly than swarms with initially 
smaller \nbodies\ (Table~\ref{tab: perc}).  For this set of 
calculations, the fraction of systems with no remaining tracers 
after 150--300 days is 28\% ($r_0$ = 145~km), 32\% ($r_0$ = 185~km), 
and 41\% ($r_0$ = 230~km).  Coupled with our result that the final 
mass of a Charon analog correlates with $r_0$, systems with a massive 
Charon analog are less likely to retain their tracers after 150--300 days.

\begin{figure}
\begin{center}
\includegraphics[width=5.0in]{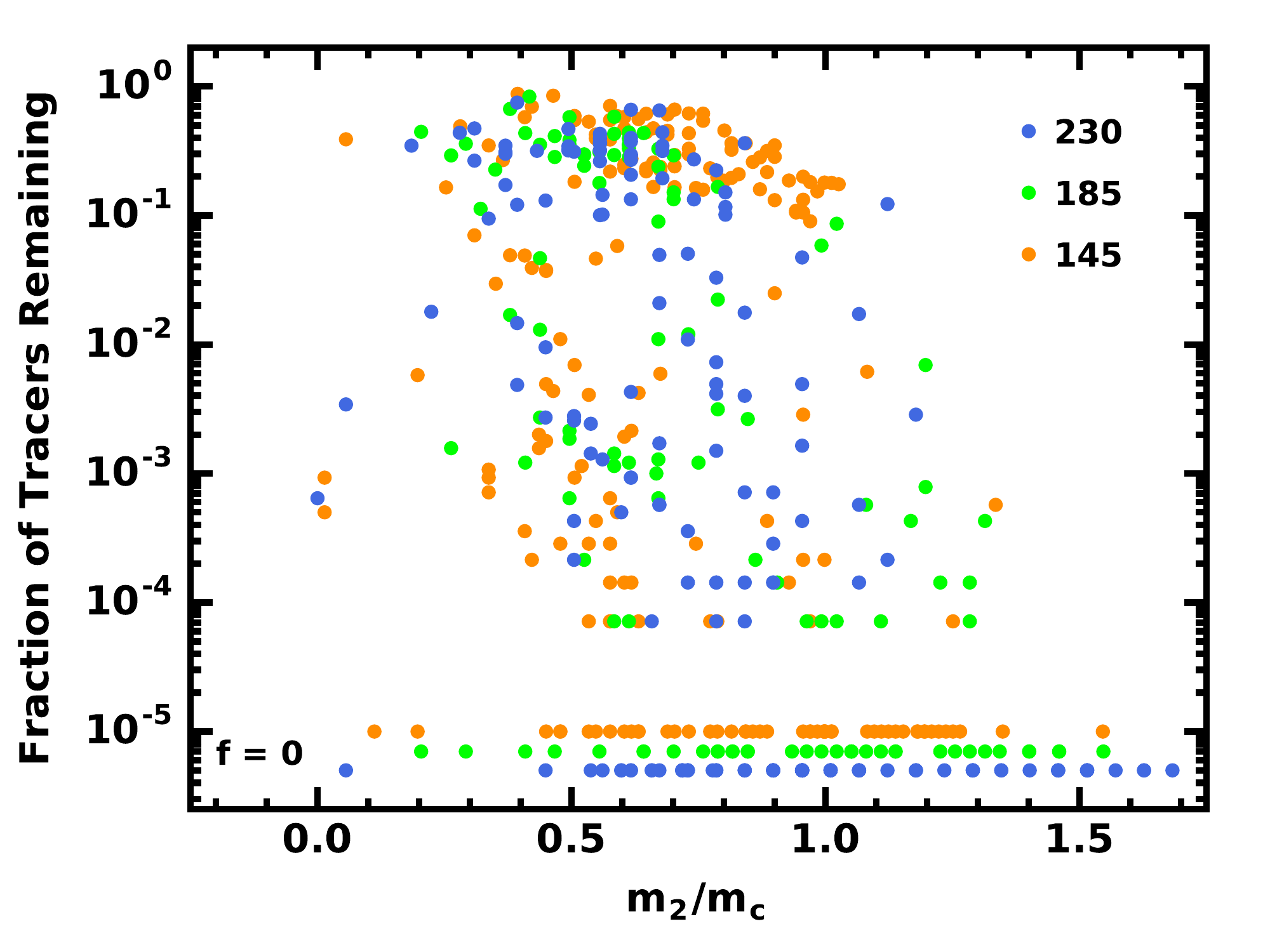}
\vskip -2ex
\caption{\label{fig: trace2}
As in Fig.~\ref{fig: trace1} with color-coding by $r_0$ the initial radius
(in km) of \nbodies\ in the swarm. Systems starting with larger \nbodies\ lose 
more tracers in 150--300~d.
}
\end{center}
\end{figure}

Fig.~\ref{fig: trace2} illustrates the trend of $f_t$ as a function 
of $\mtwo$ and $r_0$. The vast majority of the systems with 
$\mtwo\ \gtrsim$ 0.7~\mc\ and $f_t \gtrsim$ 0.1 have $r_0$ = 145~km. 
Only a few have $r_0$ = 185~km or 230~km. Among systems with 
$\mtwo\ \gtrsim$ 0.7~\mc\ and $f_t \lesssim$ 0.01, most have 
$r_0$ = 185~km or $r_0$ = 230~km (Table~\ref{tab: perc}). 

Curiously, the fraction of systems with $f_t \approx 10^{-4} - 10^{-1}$
is fairly independent of $r_0$ (Table~\ref{tab: perc}). Despite the 
large number of systems with $f_t$ = 0 and $f_t \gtrsim$ 0.1, only
$\sim$ 7\% of systems have $f_t$ = 0.01--0.1; another 10\% have 
$f_t$ = 0.001--0.01 and 15\% have $f_t = 10^{-4}$ to $10^{-3}$. 
As with tracer retention as a function of the physical properties 
of the Charon analog, this behavior -- where the systems tend to 
have $f_t \gtrsim$ 0.1 or $f_t$ = 0 -- illustrates the effectiveness
of scattered \nbodies\ in removing tracers from the system.

The statistical tests generally confirm these conclusions. All of
the tests measure a small probability for the lack of a correlation
between $r_0$ and the fraction of tracers remaining after 150--300~d
of evolution (Table~\ref{tab: corr}). For the Pearson and Kendall tests,
the correlation between $r_0$ and $f_t$ is significant at the 
3$\sigma$ or 4$\sigma$ level. The Spearman rank-order test measures 
a probability of zero that $r_0$ and $f_t$ are uncorrelated.  These 
results are mainly due to the relation between $\mtwo$ and $r_0$, 
where systems with larger planetesimals generally produce a more
massive Charon analog (\S\ref{sec: res1}). 

Among calculations where the evolution leads to the ejection of most 
tracers and the merger of the Charon analog with Pluto, there is no 
obvious preference for the initial radius of planetesimals. This 
conclusion is based on small-number statistics. Of the 13 calculations 
with $\mtwo\ \lesssim$ 0.2~\mc, seven have $f_t \lesssim 10^{-3}$ 
at 150--30~d.  Although four of the seven have $r_0$ = 145~km, one 
has $r_0$ = 185~km, and two have $r_0$ = 230~km. 

\begin{figure}
\begin{center}
\includegraphics[width=5.0in]{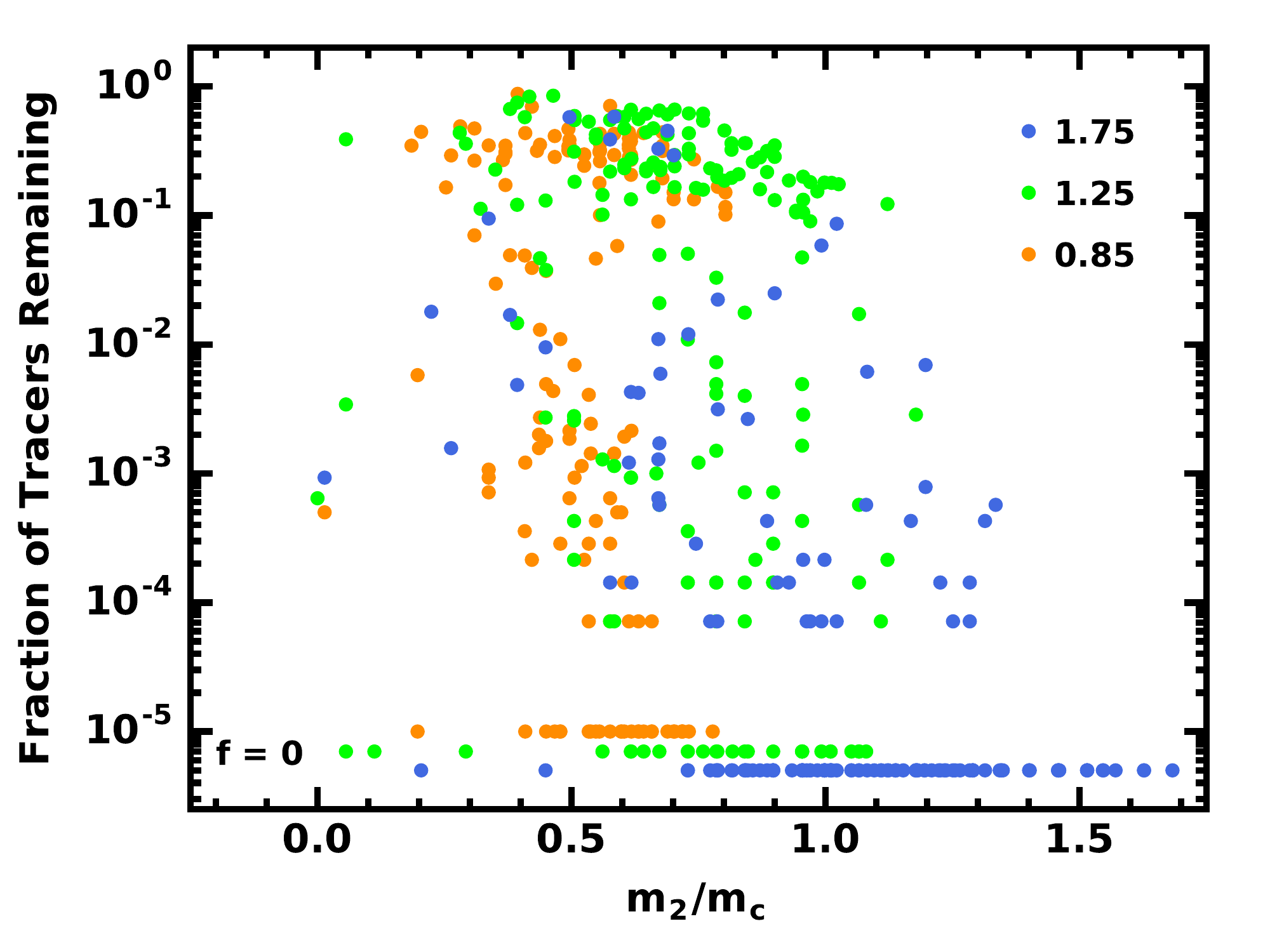}
\vskip -2ex
\caption{\label{fig: trace3}
As in Fig.~\ref{fig: trace1} with color-coding by $M_0$ the initial mass
(in units of \mc) of \nbodies\ in the swarm. Systems starting with more 
mass in \nbodies\ lose more tracers in 150--300~d.
}
\end{center}
\end{figure}

Tracer retention also depends on the initial mass of the swarm 
(Fig.~\ref{fig: trace3}). Swarms with $M_0$ = 0.85~\mc\ cannot produce a
Charon analog with the mass of Charon. Despite having a lower mass
Charon analog, these systems may still lose nearly all of their tracers 
after 150--300~d. Although 47\% of the evolutionary calculations retain 
at least 10\% of their tracers, 13\% lose 99\% to 99.9\% and another
14\% lose 99.9\% to 99.99\% (Table~\ref{tab: perc}). Roughly 20\% lose 
all of their tracers. The fraction of lost tracers loosely correlates 
with the final mass of the Charon analog: systems with $\mtwo\ \lesssim$ 
0.7~\mc\ are more likely to retain a substantial population of tracers than 
those with more massive Charon analogs.

When $M_0$ = 1.25~\mc\ (Fig.~\ref{fig: trace3}, green points), dynamical 
evolution generates a massive \nbody\ with $\mtwo\ \gtrsim$ 0.95~\mc\ (\mc)
roughly 20\% (10\%) of the time. Compared to calculations where 
$\mtwo\ \lesssim$ 0.95~\mc, these systems retain fewer tracers 
($f_t \lesssim$ 0.1) than those with a smaller massive \nbody\ 
($f_t \lesssim$ 0.8--0.9). Independent of the final mass of the Charon 
analog, these calculations retain fewer tracers overall: 41\% lose at 
least 99\% of their initial complement of tracers and 16\% lose all of 
their tracers (Table~\ref{tab: perc}).

Systems with $M_0$ = 1.75~\mc\ have an even harder time retaining tracers 
for 150--300~d (Fig.~\ref{fig: trace3}, blue points). Very few of these 
systems retain their tracers (Table~\ref{tab: perc}); 66\% lose all of them.
Unlike calculations with lower masses, these almost always generate a 
Charon-mass (or larger) satellite. In contrast with the lower $M_0$ 
trials studied here, the lack of tracers is more a function of the 
final mass of the satellite than the initial mass of the swarm.

The statistical tests indicate that the initial mass of the swarm is more 
important in setting $f_t$ than the initial radius of the planetesimals 
(Table~\ref{tab: corr}). Probabilities that $M_0$ and $f_t$ are 
uncorrelated range from $2 \times 10^{-12}$ for the Pearson test to
$6 \times 10^{-30}$ for Kendall's $\tau$ to zero for the Spearman 
rank-order test. These probabilities are comparable to those derived for
the correlation between $f_t$ and $\mtwo$. This result is not surprising:
the initial mass of the swarm is much more important in setting the final 
mass of the Charon analog than the initial radius of planetesimals.

To explore the long-term behavior of the Charon analog and any surrounding 
tracers, we followed the evolution of selected systems for 10--100~yr. 
During this period, the physical properties of the Charon analogs change 
little. Their masses remain constant. Their orbital semimajor axes and 
eccentricities settle onto constant values that differ little from those
at 1~yr. Thus, the binary planet composed of Pluto and a Charon analog is
stable.

\begin{figure}
\begin{center}
\includegraphics[width=5.0in]{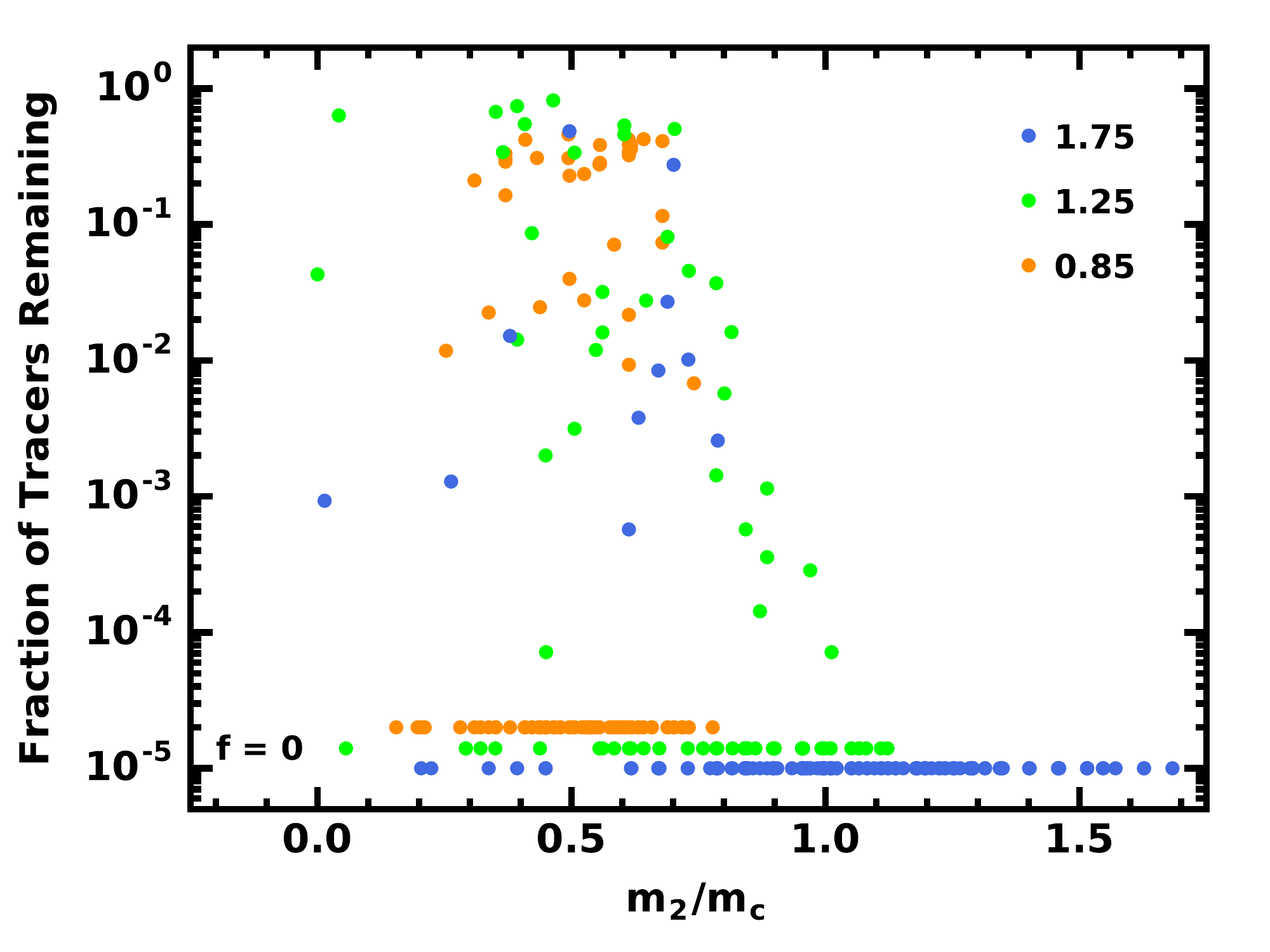}
\vskip -2ex
\caption{\label{fig: trace4}
As in Fig.~\ref{fig: trace3} for a subset of systems after 10--100~yr of 
dynamical evolution. For swarms of \nbodies\ that generate a massive Charon
analog, nearly all tracers are lost within 100~yr.
}
\end{center}
\end{figure}

As the Charon analog reaches a stable state, the evolution in the tracer
population is dramatic (Fig.~\ref{fig: trace4}). All systems with a 
massive Charon analog, $\mtwo\ \gtrsim$ 0.9~\mc, lose 99.9\% of their
tracers. When $\mtwo\ \gtrsim$ 0.95~\mc, only two of 63 calculations retain
any tracers, one with four and another with one. Both of these calculations
have $M_0$ = 1.25~\mc. None of the calculations with $M_0$ = 1.75~\mc\ that
produce a Charon analog with $\mtwo\ \gtrsim$ 0.9~\mc\ retain any tracers 
after 10--100~yr of dynamical evolution.

The ability to retain tracers after 10--100~yr clearly correlates with the
mass of the Charon analog. In systems with $\mtwo\ \approx 0.5 ~\mc$, the 
number of tracers remaining after 10--100~yr is similar to the number at 
150--300~d.  Low mass Charon analogs are not as efficient at ejecting 
leftover \nbodies; most leftovers simply collide and merge with Pluto. 

From the formation of a massive Charon analog to 10--100~yr, the timing 
of tracer ejection is random. In some systems, a few remaining low mass 
\nbodies\ make close approaches to either Pluto or the Charon analog early 
in the evolution. These are rapidly ejected and remove any remaining
tracers on their way out of the system. In other calculations, it may take
many months for Pluto or Charon to eject several left over \nbodies. 
Although all systems with a massive Charon analog have no remaining tracers
at 10--100~yr, the epoch when the last tracers are ejected depends more on
the chaotic nature of a particular system than the mass of the Charon 
analog or the time when it reaches its final mass.

Other properties of the system -- initial conditions ($e_0$, $r_0$ and 
$M_0$) and the final orbit of the Charon analog ($a_2$ and $e_2$) -- have 
little impact on the tracer population at 10--100~yr. Once the mass of 
the Charon analog is set, the tracers `forget' the initial conditions 
and respond only to the dynamics created by the Charon analog. When the
Charon analog is massive, 
(i) it creates a tracer population with somewhat larger semimajor axes
$a_t$ and eccentricities $e_t$ than lower mass Charon analogs,
(ii) it ejects more left over \nbodies\ through the population of 
tracers, and
(iii) its stronger gravity more strongly destabilizes tracers with large
$e_t$. As a results, tracers do not survive in a system with a massive 
Charon.

To understand whether this dynamical evolution of the tracers is characteristic,
we performed several calculations with a simple collisional damping model. For
each tracer with semimajor axis $a$, the change in $e$ and $\imath$ is
\begin{equation}
{de \over dt} = {e \over t_{damp}}
\end{equation}
and 
\begin{equation}
{d\imath \over dt} = {\imath \over t_{damp}} ~ ,
\end{equation}
where $t_{damp}$ is a damping time appropriate for collisions among particles with
radii $r$ and total mass $M_t$ \citep{bk2015b}:
\begin{equation}
t_{damp} = 1250 \left ( {a \over \rp} \right )^{3/2}
\left ( { r \over {\rm 1~cm} } \right )
\left ( { 10^{22}~{\rm g} \over M_t } \right ) ~ {\rm sec} ~ .
\end{equation}
If this damping can overcome excitation by massive \nbodies, tracers might remain on
circular orbits far from the central binary. 

Experiments with $r \approx$ 1--100~cm and $M_t = 10^{22} - 10^{23}$~g yield similar 
results. Although some tracers manage to achieve fairly circular orbits at $a \approx$
50--60~\rp, the set of lower mass \nbodies\ sweeping through the tracers every orbit
tends to scatter a few of them to larger $a$. The longer damping times at larger $a$
preclude efficient damping; tracers at larger $a$ are then ejected. Over the course of
10--300 days, the cumulative impact of the \nbodies\ removes nearly all of the tracers
orbiting a central \pc\ binary.

\vskip 6ex
\section{DISCUSSION}
\label{sec: disc}

The calculations described here are the first to explore the growth of a Charon mass
satellite following a graze-and-merge collision between a planetary embryo and Pluto.
In systems where the debris has an initial mass $M_0$ = 0.85--1.75~\mc\ at semimajor 
axes $a \approx$ 3--11~\rp, collisional evolution almost always leads to a massive 
satellite with a significant fraction of Charon's mass.  When $M_0$ = 1.25~\mc\ 
(1.75~\mc), $\sim$ 10\% (55\%) of the calculations yield a satellite with 
$\mtwo \gtrsim$ \mc. Overall, 12\% ($M_0$ = 1.25~\mc) to 20\% ($M_0$ = 1.75~\mc) of the 
calculations produce a satellite with $\mtwo \approx$ 0.95--1.05~\mc. Given sufficient 
mass in a circum-Pluto ring of planetesimals at 3--11~\rp, massive satellite 
formation is a fairly robust outcome of a graze-and-merge impact \citep[see also][]
{canup2001,lein2010,asphaug2014}.

The approach adopted here neglects the impact of tidal stresses on the shapes
and collision outcomes of massive solids inside the Roche limit. When the solids
in these calculations venture inside the Roche limit, they probably do not have 
time to develop the expected equilibrium shapes \citep[e.g.,][]{quill2016}. Thus,
a more accurate treatment of shapes probably has little impact on the growth of
Charon mass objects. Although we confirm that no collisions occur inside the Roche 
limit, an improved treatment would begin with material inside the Roche limit 
\citep[as suggested by the SPH calculations;][]{canup2005,canup2011} and follow 
the evolution of solids inside and outside the Roche limit. Although it seems 
unlikely that including material inside the Roche limit would change results
significantly, we plan to examine this possibility in a future study.

In these calculations, we have assumed that the mass density of individual solids 
within the circum-Pluto ring of planetesimals nearly matches the mass density of Charon. 
This assumption runs counter to the SPH results of \citet{canup2005,canup2011}, who 
predicted a much smaller mass density for solids within the debris from a 
graze-and-merge collision between two partially differentiated embryos. In this 
scenario, most of the rocky material within the impactor merges with Pluto. Any
circum-Pluto material is then mostly ice.  However, \citet{desch2015} and 
\citet{desch2017} suggest the colliding embryos probably have undifferentiated 
crusts. Graze-and-merge impacts then leave behind substantial rocky material in 
a circum-Pluto disk.

The degree of differentiation of colliding embryos depends on their formation 
and evolution histories.  In a protosolar disk where embryos grow rapidly from 
pebble accretion \citep[e.g.,][]{chambers2016,johansen2017}, large solids may
retain significant amounts of short-lived radioactivities, which serve as a 
major heat source following formation. Analysis of data from the \nh\ mission
appears to rule out this possibility \citep{mckinnon2017}. The slower growth of
embryos from 1--10~km planetesimals \citep[e.g.][]{kb2008,kb2010,kb2012} allows 
heating only from longer-lived radioactivities \citep[e.g.,][]{malamud2015} and
may be more consistent with the partially differentiated objects required in 
hit-and-run or graze-and-merge impacts. Given the uncertainty in formation 
histories, it is worthwhile to consider in more detail how the outcomes of 
giant impacts depend on the formation mechanism.

\subsection{Growth Time and Long-term Dynamical Evolution of Charon Analogs}
\label{sec: disc-dyn}

Starting with swarms of 145--230~km planetesimals orbiting Pluto at 3--11~\rp, 
Charon analogs grow rapidly. Within a day, more than half of the planetesimals 
have merged into a larger object. After 10~days, only a Charon analog and several 
low mass objects remain. By 100 days, the Charon analog has usually accreted or 
ejected all of the leftovers.

Although the growth of Charon analogs seems rapid, the 100 day time scale is 
in line with theoretical expectations \citep[see also][for a discussion of 
SPH calculations of satellite evolution after a giant impact]{arakawa2019}. 
At $a \approx$ 50--60~\rp, it takes $\sim$ 
0.1 $(M_0 / {\rm 3 ~ \times ~ 10^{20}~g)}$~Myr to convert an initial mass $M_0$ 
of small planetesimals orbiting Pluto into several large objects that contain 
most of the initial mass of the swarm \citep{kb2014b}. For the initial masses 
considered here, $M_0 \approx 3 \times 10^{24}$~g, the growth time is $10^4$ 
times smaller, only 10~yr.  In addition to the scaling between growth time and 
the initial mass of the swarm, ensembles of solids evolve on a time scale 
proportional to the orbital period \citep[e.g.,][]{lissauer1987,lissauer1993,
gold2004,kb2006}. Scaling results at 50--60~\rp\ to 5--6~\rp\ yields an expected 
growth time of $\sim$ 100~days, which agrees with the typical growth time of 
the Charon analogs described here.

The final orbits of Charon analogs in these calculations are similar to those of 
the Charon-mass survivors of hit-and-run collisions \citep{canup2005,canup2011}. 
Thus, the subsequent tidal evolution of Pluto and Charon should be fairly independent 
of the formation mechanism.  With $a_C \approx$ 5--6~\rp\ and $e_C \approx$ 0.1--0.4, 
it takes $\sim$ 1~Myr for tidal forces between Pluto and Charon to circularize the 
orbit and to migrate Charon to its current semimajor axis \citep{farinella1979,dobro1997,
peale1999,cheng2014a,barr2015}. These analyses assume that the immediate environment 
of \pc\ has 
no other large objects with significant gravity. For the outcomes of graze-and-merge 
scenarios considered here, the \pc\ binary clears out the inner 25--50~\rp\ of the 
binary within 1--100~yr. In systems where the leftover solids are much smaller than 
150--200~km, mass removal might take somewhat longer, but the clearing time should 
still be much smaller than 1~Myr \citep[see also][]{winter2010,giuliatti2013,
giuliatti2014,giuliatti2015,kb2019a}. 

\subsection{Long-term Thermal Evolution of Charon Analogs}
\label{sec: disc-the}

Although examining the long-term internal evolution of Charon requires detailed 
structural calculations with state-of-the-art equations-of-state 
\citep[e.g.,][]{robuchon2011,malamud2013,desch2015,malamud2015,mckinnon2016,
hammond2016,desch2017,mckinnon2017,bierson2018}, it is worth considering 
whether the rapid formation time might establish somewhat different initial 
conditions for this evolution.  Once it reaches its final mass, Charon has 
a gravitational binding energy 
\begin{equation}
E \approx {3 G \mc^2 \over 5 \rc } ~ ,
\label{eq: Ebind}
\end{equation}
where we assume for simplicity a uniform sphere of constant density. With
$\mc = 1.6 \times 10^{24}$~g and \rc\ = 606~km, $E \approx 2 \times 10^{33}$~erg.
Setting the internal energy $U = NkT = E/2$, where $N$ is the number of molecules
in Charon, $k$ is Boltzman's constant, and $T$ is the internal temperature:
\begin{equation}
T = { E \over 3 N k } 
\label{eq: Eint}
\end{equation}
Making the simple assumption that Charon is 100\% water ice, $N = \mc / 18 m_H$, 
where $m_H$ is the mass of a hydrogen atom. Then, $T \approx$ 90 K. A Charon 
composed of 50\% water ice and 50\% silicon has $T \approx$ 70 K. 

It is unlikely that Charon analogs formed during a graze-and-merge impact radiate 
the gravitational binding energy efficiently.  Adopting an effective temperature 
$T_e$ = 90~K (70~K), Charon has a luminosity 
$L_C \approx 2 \times 10^{20}$~\ergs\ ($L_C \approx 6 \times 10^{19}$~\ergs).
The time scale required to radiate $\sim$ 50\% of the binding energy is $\sim$ 
0.2~Myr (0.5~Myr). Thus, Charon retains the heat of formation over a time 
comparable to the tidal expansion time scale.

Despite the relatively slow cooling compared to planetary embryos grown from 
km-sized or larger planetesimals, the $T$ = 70--90~K derived above is similar
to the initial temperature adopted for several detailed evolutionary calculations,
$T \approx$ 100~K \citep{malamud2015} and $T$ = 60--100~K \citep{desch2017}. In
these studies, radioactive decay of $^{235}$U and other long-lived unstable 
elements powers a gradual rise in the core temperature $T_c$, which reaches 
$T_c \approx$ 500--1000~K $\sim$ 1~Gyr after the impact. This core temperature
is sufficient to melt water ice and produce a differentiated planet \citep[see 
also][] {mckinnon2017}. Although the core temperature is sensitive to composition 
and the details of the equation-of-state, it is rather insensitive to $E$. 

Aside from radioactive decay, other sources of heating are also negligible. The 
energy in the orbit, $E_o \approx G \mp \mc / a \sim 2 \times 10^{33}$~erg, 
is comparable to the gravitational binding energy of Charon. Unless the
rotational periods of Pluto and Charon are much less than 1 day, the available 
rotational energy is roughly equal to the orbital energy. Thus, tidal expansion 
has modest impact on the thermal energy of either Pluto or Charon \citep{dobro1997,
barr2015}. 
Collisions with 100--200~km Kuiper belt objects only add significant energy 
if the impacts occur at velocities much larger than the orbital velocity of 
\pc\ around the Sun, which is very unlikely. Overall, the rapid formation of 
Charon in the graze-and-merge scenario does not change expectations for its
thermal evolution.

\subsection{Future Prospects: Survival of the Circumbinary Satellite System}
\label{sec: disc-sat}

However Charon forms, the dynamical evolution of massless tracers described 
in \S\ref{sec: res2} places strong constraints on the aftermath of hit-and-run 
and graze-and-merge impacts. In either scenario, some debris orbits the system 
barycenter well outside Charon's orbit.  For typical surface densities within 
the debris at 30--50~\rp, collisional damping rapidly circularizes the orbits 
of small particles with radii $r \lesssim$ 0.1~km \citep{bk2015b}. 
On time scales of $10^4 - 10^5$ yr, collisional evolution can then transform 
swarms of small particles into an ensemble of 5--20~km satellites \citep{kb2014b}. 
However, if the debris contains several massive particles with radii 
$r \gtrsim$ 150--200~km, dynamical ejection of only one or two of these 
objects by the \pc\ binary is sufficient to disrupt the entire swarm of 
circumbinary particles.  

In the calculations described here, the massless tracers serve as proxies for
small particles produced in the giant impact or from the collisions of massive 
\nbodies\ as they merge to produce a Charon analog orbiting Pluto.  If the 
small particles constitute a fraction $f_s$ of the initial mass in solids, 
the total mass in small particles at the end of a calculation is
$M_s \approx f_t f_s M_0$, where $f_t$ is the fraction of tracers remaining
after 10--100~yr of evolution. Recalling $M_{SNKH}$ as the total mass of the
known circumbinary satellites, setting $M_s \gtrsim M_{SNKH}$ allows the 
surviving tracers to have sufficient mass to form Styx, Nix, Kerberos, and
Hydra on a reasonable time scale. With $M_0 \approx$ 1--2~$ \mc\ $
$\lesssim 3 \times 10^{24}$~g and $M_{SNKH} \lesssim 10^{20}$~g, 
we require $f_t f_s \gtrsim 10^{-4}$.

Our calculations suggest this constraint on $f_t f_s$ is not achievable with
the initial conditions considered here. From SPH simulations of giant impacts,
having more than 1\% of the debris in the form of small particles seems unlikely
\citep[e.g.,][]{canup2005,canup2011,arakawa2019}. Debris from the collisions of
massive icy \nbodies\ is also unlikely to exceed 1\% (\S\ref{sec: sims}).  For 
swarms of massive \nbodies\ capable of producing a massive Charon analog, 
$f_t \lesssim 10^{-4}$. Thus, the mass remaining in small particles is not
large enough to produce the current circumbinary satellites.  

Aside from the limited available mass, any surviving tracers are well 
beyond the orbits of the circumbinary satellites. Scattering by massive 
\nbodies\ places tracers on eccentric orbits with typical semimajor axes 
$a_t \approx$ 125--175~\rp, outside the current orbit of Hydra, 
$a_H \approx$ 55~\rp. On time scales of $10^1 - 10^6$ yr, massive 
collections of small particles with radii $r \approx$ 1~m to 1~km can 
spread inward from $\sim$ 150~\rp\ to the orbit of Hydra \citep{bk2015b}. 
If sufficient numbers of tracers survive the formation of a Charon analog,
the current ensemble of circumbinary satellites might form in a spreading
disk of debris.

Growing Charon within a swarm of lower mass planetesimals might allow
enough massless tracers to survive the ejections of any leftovers.  For 
material orbiting Pluto at 50-200~\rp, the orbital velocity is $\sim$ 
0.07--0.14~\kms.  With escape velocities $v_e \approx$ 0.14--0.23~\kms, 
massive planetesimals with $r$ = 145--230~km stir up nearby tracers to 
velocities larger than the local escape velocity. Thus, tracers are ejected. 
Lower mass planetesimals with $r$ = 10--20~km have a factor of ten smaller 
escape velocities and therefore stir tracers to velocities $\sim$ 10\% of 
the local escape velocity. Small particles with these velocities might 
survive collisional evolution within circumbinary rings \citep{kb2014b,
walsh2015,bk2015b}. 

Following the collisional evolution of swarms of particles with $r \lesssim$ 
10--20~km at semimajor axes $a \approx$ 5--20~\rp\ requires an accurate 
treatment of fragmentation \citep[e.g.,][]{kb2014b}. For typical velocities 
of 0.05--0.10~\kms\ at $a \approx$ 5--10~\rp, collisions of equal mass 
10--20~km objects eject 5\% to 10\% of the combined mass in small particles. 
Unlike the calculations of 145--230~km objects here, the large amount of 
debris generated in collisions of much smaller objects may change the growth 
of Charon analogs and the ejection of small particles by surviving \nbodies. 
In calculations of icy objects with fragmentation, leftover planetesimals 
are often small \citep[e.g.,][]{kb2008,kb2010,kb2014b}.  Dynamical friction 
between large objects and collisional debris maintains large objects on 
fairly circular orbits. Thus, there is some chance that starting with a 
swarm of much lower mass planetesimals might allow the survival of a 
circumbinary swarm of debris at 30--60~\rp.  We plan to consider these kinds
of calculations in a future study.

If small particles at 25--100~\rp\ can survive the immediate aftermath of a 
graze-and-merge collision, this material must somehow endure the tidal evolution 
of the binary \citep[e.g.,][]{ward2006,lith2008a,cheng2014b,bk2015b,smullen2017,woo2018}.  
As the binary circularizes and expand, orbits with small integer ratios of the 
binary orbital period become unstable. As an example, the 5:1 resonance passes
through material orbiting at 30~\rp\ (just inside the current orbit of Styx) 
when the central binary has a period of 4.5 days. When the expansion is rapid, 
some material survives on high eccentricity orbits. However, current models 
suggest circularization and expansion are rather slow. Few solids survive this
expansion \citep[see also][]{walsh2015}.

Collisional damping allows solids orbiting at 30--60~\rp\ to survive tidal 
evolution of the central binary \citep{bk2015b}. When a resonance encounters a 
ring of m-sized or somewhat larger particles, the eccentricities of the particles
begin to grow. However, collisions among the swarm damp the eccentricity as fast
as the resonance excites it. Damped particles are then transported out with the
resonance. Although some particles are lost, most survive the expansion and lie
within resonances as the expansion ends.

This mechanism offers a way to produce circumbinary satellites after tidal
expansion following either a graze-and-merge or a hit-and-run impact. If the 
debris at 30--60~\rp\ is composed of km-sized or smaller particles, collisions
will circularize the debris on time scales short compared to the tidal expansion
time scale \citep{kb2014b,bk2015b}. As the binary expands, these particles become
trapped in resonance. Once tidal expansion ceases, collisional processes convert
the survivors into a few small satellites near resonance. The main challenge is
for the solids to avoid growing into satellites before tidal expansion ends. We
plan to describe numerical simulations of this process in a future paper.

\vskip 6ex
\section{SUMMARY}
\label{sec: summary}

We consider the evolution of massive disks of planetesimals orbiting a planet
with mass similar to Pluto. Within a few weeks, 145--230~km survivors of a 
graze-and-merge impact grow into a Charon analog with an orbit -- $a_C \approx$ 
5--6~\rp\ and $e_C \approx$ 0.1--0.3 -- similar to the orbits of Charons that
survive a hit-and-run collision and remain bound to Pluto. In a typical calculation,
the Charon analog contains $\sim$ 60\% of the initial mass in large objects. 
Pluto typically accretes roughly 25\% of the initial mass and ejects the rest. 

When a swarm of planetesimals produces a Charon analog with $m_2 \gtrsim$ 0.9~\mc, 
nearly all massless tracers are ejected on time scales of months to decades. Initially, 
the growth of the Charon analog places tracers on eccentric orbits with $a \approx$ 
50--250~\rp. Although collisional damping would probably circularize the tracers 
on long time scales \citep[e.g.,][]{kb2014b,bk2015b}, the central binary ejects 
several leftover massive planetesimals through the tracers. These ejections 
disrupt tracer orbits; eventually all of the tracers are also ejected. 
Thus, there is no circumbinary disk of solids in which to grow satellites with 
properties similar to the known small satellites. 

A simple dynamical analysis suggests that solids at $a \gtrsim$ 30~\rp\ would 
survive the passage and subsequent ejection of less massive objects with 
$r \lesssim$ 10--20~km. If Charon grows efficiently within a circum-Pluto 
debris disk and ejects leftovers with radii no larger than 10--20~km, it might 
then be possible to retain a circumbinary disk of solids and form small 
satellites at 30--60~\rp.
We plan to test this scenario in a future study.

\vskip 6ex

We acknowledge generous allotments of computer time on the NASA `discover' cluster.
Advice and comments from M. Geller greatly improved our presentation.  We thank 
the referee for a timely and useful report.  Portions of this project were supported 
by the {\it NASA } {\it Outer Planets} and {\it Emerging Worlds} programs through 
grants NNX11AM37G and NNX17AE24G.

Binary output files from the simulations and C programs capable of reading the 
binary files are available at a publicly accessible repository 
(https://hive.utah.edu/) with digital object identifier
https://doi.org/10.7278/S50D-EFCY-ZC00.

%\bibliography{sfpl}
\bibliography{ms.bbl}

\end{document}

%% file: defs.tex
\def\nbody{$n$-body}
\def\nbodies{$n$-bodies}
\def\deg{\ifmmode {^\circ}\else {$^\circ$}\fi}
\def\degree{\ifmmode {^\circ}\else {$^\circ$}\fi}
\def\mum{\ifmmode {\rm \,\mu {\rm m}}\else $\rm \,\mu {\rm m}$\fi}
\def\arcsec{\ifmmode ^{\prime \prime}\else $^{\prime \prime}$\fi}

\def\inch{\ifmmode ^{\prime \prime}\else $^{\prime \prime}$\fi}
\def\gs{\ifmmode {{\rm g~s^{-1}}}\else ${\rm g~s^{-1}}$\fi}
\def\msunyr{\ifmmode {M_{\odot}~{\rm yr^{-1}}}\else $M_{\odot}~{\rm yr^{-1}}$\fi}
\def\msun{\ifmmode {M_{\odot}}\else $M_{\odot}$\fi}
\def\rsun{\ifmmode {R_{\odot}}\else $R_{\odot}$\fi}
\def\lsun{\ifmmode {L_{\odot}}\else $L_{\odot}$\fi}
\def\mstar{\ifmmode {M_{\star}}\else $M_{\star}$\fi}
\def\rstar{\ifmmode {R_{\star}}\else $R_{\star}$\fi}
\def\tstar{\ifmmode {T_{\star}}\else $T_{\star}$\fi}
\def\lstar{\ifmmode {L_{\star}}\else $L_{\star}$\fi}
\def\mwd{\ifmmode {M_{wd}}\else $M_{wd}$\fi}
\def\rwd{\ifmmode {R_{wd}}\else $R_{wd}$\fi}
\def\twd{\ifmmode {T_{wd}}\else $T_{wd}$\fi}
\def\lwd{\ifmmode {L_{wd}}\else $L_{wd}$\fi}
\def\md{\ifmmode {M_d}\else $M_d$\fi}
\def\ld{\ifmmode {L_d}\else $L_d$\fi}
\def\ad{\ifmmode A_d\else $A_d$\fi}
\def\ldlwd{\ifmmode L_d / L_{wd}\else $L_d / L_{wd}$\fi}
\def\ldlstar{\ifmmode L_d / L_\star\else $L_d / L_{\star}$\fi}
\def\rearth{\ifmmode {\rm R_{\oplus}}\else $\rm R_{\oplus}$\fi}
\def\mearth{\ifmmode {\rm M_{\oplus}}\else $\rm M_{\oplus}$\fi}
\def\qdstar{\ifmmode Q_D^\star\else $Q_D^\star$\fi}
\def\vsqd{\ifmmode v^2 / Q_D^\star\else $v^2 / Q_D^\star$\fi}
\def\kms{\ifmmode {\rm km~s^{-1}}\else $\rm km~s^{-1}$\fi}
\def\ms{\ifmmode {\rm m~s^{-1}}\else $\rm m~s^{-1}$\fi}
\def\vrel{\ifmmode v_{rel}\else $v_{rel}$\fi}
\def\mdot{\ifmmode \dot{M}\else $\dot{M}$\fi}
\def\mdotz{\ifmmode \dot{M}_0\else $\dot{M}_0$\fi}
\def\mesc{\ifmmode m_{esc}\else $m_{esc}$\fi}
\def\rmin{\ifmmode r_{min}\else $r_{min}$\fi}
\def\rmax{\ifmmode r_{max}\else $r_{max}$\fi}
\def\xmax{\ifmmode x_{max}\else $x_{max}$\fi}
\def\mmin{\ifmmode m_{min}\else $m_{min}$\fi}
\def\mmax{\ifmmode m_{max}\else $m_{max}$\fi}
\def\rmind{\ifmmode r_{min,d}\else $r_{min,d}$\fi}
\def\rmaxd{\ifmmode r_{max,d}\else $r_{max,d}$\fi}
\def\mmaxd{\ifmmode m_{max,d}\else $m_{max,d}$\fi}
\def\vrad{\ifmmode v_{rad}\else $v_{rad}$\fi}
\def\qz{\ifmmode q_{0}\else $q_{0}$\fi}
\def\qi{\ifmmode q_{i}\else $q_{i}$\fi}
\def\ql{\ifmmode q_{l}\else $q_{l}$\fi}
\def\qs{\ifmmode q_{s}\else $q_{s}$\fi}
\def\vhill{\ifmmode v_H\else $r_H$\fi}
\def\rhill{\ifmmode r_H\else $r_H$\fi}
\def\Rhill{\ifmmode R_H\else $R_H$\fi}
\def\rbrk{\ifmmode r_{brk}\else $r_{brk}$\fi}
\def\rdamp{\ifmmode r_{damp}\else $r_{damp}$\fi}
\def\rin{\ifmmode r_{in}\else $r_{in}$\fi}
\def\rout{\ifmmode r_{out}\else $r_{out}$\fi}
\def\tin{\ifmmode t_{in}\else $t_{in}$\fi}
\def\tout{\ifmmode t_{out}\else $t_{out}$\fi}
\def\ain{\ifmmode a_{in}\else $a_{in}$\fi}
\def\aout{\ifmmode a_{out}\else $a_{out}$\fi}
\def\r0{\ifmmode r_{0}\else $r_{0}$\fi}
\def\R0{\ifmmode R_{0}\else $R_{0}$\fi}
\def\m0{\ifmmode m_{0}\else $m_{0}$\fi}
\def\mone{\ifmmode m_{1}\else $m_{1}$\fi}
\def\mtwo{\ifmmode m_{2}\else $m_{2}$\fi}
\def\atwo{\ifmmode a_{2}\else $a_{2}$\fi}
\def\etwo{\ifmmode e_{2}\else $e_{2}$\fi}
\def\mf{\ifmmode m_{f}\else $m_{f}$\fi}
\def\af{\ifmmode a_{f}\else $a_{f}$\fi}
\def\ef{\ifmmode e_{f}\else $e_{f}$\fi}
\def\M0{\ifmmode M_{0}\else $M_{0}$\fi}
\def\amax{\ifmmode a_{max}\else $a_{max}$\fi}
\def\a0{\ifmmode a_{0}\else $a_{0}$\fi}
\def\e0{\ifmmode e_{0}\else $e_{0}$\fi}
\def\v0{\ifmmode v_{0}\else $v_{0}$\fi}
\def\xm{\ifmmode x_{m}\else $x_{m}$\fi}
\def\sigz{\ifmmode \Sigma_0\else $\Sigma_0$\fi}
\def\ergg{\ifmmode {\rm erg~g^{-1}}\else ${\rm erg~g^{-1}}$\fi}
\def\ergs{\ifmmode {\rm erg~s^{-1}}\else ${\rm erg~s^{-1}}$\fi}
\def\gyr{\ifmmode {\rm g~yr^{-1}}\else ${\rm g~yr^{-1}}$\fi}
\def\cms{\ifmmode {\rm cm~s^{-1}}\else ${\rm cm~s^{-1}}$\fi}
\def\gcms{\ifmmode {\rm g~cm^{-2}}\else $\rm g~cm^{-2}$\fi}
\def\gcmc{\ifmmode {\rm g~cm^{-3}}\else $\rm g~cm^{-3}$\fi}
\def\atil{\ifmmode {\tilde{a}}\else $\tilde{a}$\fi}
\def\ttil{\ifmmode {\tilde{t}}\else $\tilde{t}$\fi}
\def\sqrttt{\ifmmode {\tilde{t}^{1/2}}\else $\tilde{t}^{1/2}$\fi}

\def\orch{{\it Orchestra}}
\def\nh{{\it New Horizons}}
\def\pc{Pluto--Charon}
\def\mp{\ifmmode m_P\else $m_P$\fi}
\def\mc{\ifmmode m_C\else $m_C$\fi}
\def\mh{\ifmmode m_H\else $m_H$\fi}
\def\mk{\ifmmode m_K\else $m_K$\fi}
\def\ms{\ifmmode m_S\else $m_S$\fi}
\def\mn{\ifmmode m_N\else $m_N$\fi}
\def\rp{\ifmmode r_P\else $r_P$\fi}
\def\rc{\ifmmode r_C\else $r_C$\fi}
\def\apc{\ifmmode a_{PC}\else $a_{PC}$\fi}
\def\mpc{\ifmmode m_{PC}\else $m_{PC}$\fi}
\def\epc{\ifmmode e_{PC}\else $e_{PC}$\fi}